\documentclass[twocolumn,showpacs,preprintnumbers,amsmath,amssymb, floatfix]{revtex4}

\usepackage{subfigure} 
\usepackage{graphicx}
\usepackage{dcolumn}
\usepackage{bm}

\begin{document}


\title{Structure and dynamics of colloidal depletion gels: \\coincidence of
transitions and heterogeneity
}

\author{C. J. Dibble}
\author{M. Kogan}%
\author{M. J. Solomon}%
 \email{mjsolo@umich.edu}
\affiliation{%
University of Michigan, Ann Arbor, Michigan 48109-2136
}%


\date{\today}

\begin{abstract}
Transitions in structural heterogeneity of colloidal depletion gels formed through short-range attractive interactions are correlated with their dynamical arrest.  The system is a density and refractive index matched suspension of 0.20 volume fraction poly(methyl methacyrlate) colloids with the non-adsorbing depletant polystyrene added at a size ratio of depletant to colloid of 0.043.  As the strength of the short-range attractive interaction is increased, clusters become increasingly structurally heterogeneous, as characterized by number-density fluctuations, and dynamically immobilized, as characterized by the single-particle mean-squared displacement.  The number of free colloids in the suspension also progressively declines.  As an immobile cluster to gel transition is traversed, structural heterogeneity abruptly decreases.  Simultaneously, the mean single-particle dynamics saturates at a localization length on the order of the short-range attractive potential range.   Both immobile cluster and gel regimes show dynamical heterogeneity.  Non-Gaussian distributions of single particle displacements reveal enhanced populations of dynamical trajectories localized on two different length scales.  Similar dependencies of number density fluctuations, free particle number and dynamical length scales on the order of the range of short-range attraction suggests a collective structural origin of dynamic heterogeneity in colloidal gels.  

\end{abstract}

\maketitle

\section{\label{S:intro}Introduction}
Improved understanding of the origin and consequences of slow
dynamics in disordered colloidal suspensions impacts their application in
both established \cite{SoftFoodReview2005NatMat} and emerging \cite{Shanbhag2005Small}
technologies. Slow dynamics in colloidal gels and glasses can arise under a
broad range of suspension packing fractions, potential interactions, and
structures.  At dense volume fractions, without any clear structural
transition \cite{Conrad2005JPCB}, colloidal hard spheres form a glass due to
caging that is characterized by dynamics \cite{Kegel2000Sci, Weeks2000Sci}
several orders of magnitude slower than its liquid precursor.  Gels
comprised of colloids interacting through short-range attractive forces
share this dynamic arrest \cite{Segre2001PRL}; however, here the mechanistic
origin is unclear, since observed gel structures are highly variable -
ranging from the ramified fractal clusters of dilute gels formed from
aggregation into deep potential wells \cite{Carpineti1992PRL} to the nearly equilibrium liquid
structures of densely packed attractive glasses \cite{Pham2002Sci}.  Understanding the
effect of manipulating control parameters on microstructure and
microdynamics is the key to deeper understanding of gel phase
behavior, rheological response and processing characteristics.

As volume fraction is increased, various phenomenological regimes
have been observed including fractal cluster gels \cite{Weitz1984PRL}, mobile
and immobile cluster phases \cite{Lu2006PRL, Ramakrishnan2005Lang, Sedgwick2004JPCM, Stradner2004Nat}, concentrated attractive gels and attractive glasses \cite{Pham2002Sci}.  Although open questions in all these areas remain, the range of intermediate particle loading is poorly
understood in particular.  In this regime, typically characterized by both intermediate \cite{Shah2003JPCM} and long-range \cite{Varadan2003Lang} structural heterogeneity, neither the micromechanical ideas based on fractal dynamics \cite{Krall1998PRL} nor the excluded volume caging mechanisms of glasses \cite{Weeks2000Sci} can be applied.

This regime of concentrated gelation exhibits complex structural
transitions \cite{Cates2004JPCM, Manley2005PRL, Verduin1995JCIS, Grant1993PRE, Elliott2003FarD}.  In addition to network gels in various degrees of arrest and compaction \cite{Lu2006PRL}, experiments have recently revealed cluster fluid phases \cite{Sedgwick2004JPCM, Lu2006PRL} and ordered tetrahedral
chains \cite{Campbell2005PRL}.  Moreover, within this intermediate volume
fraction range, clustering and gelation boundaries may lie in close
proximity to metastable liquid/crystal and liquid/liquid equilibrium phase
boundaries \cite{Pham2004PRE}.  Gel boundaries also can display a strong kinetic component
that leads to transient gelation in non-density matched suspensions \cite{Pusey1993Pa}.

Mechanistic descriptions of the dynamic transitions that accompany these
regimes include predictions of particle localization due to strong short-range interactions from direct simulation \cite{Puertas2002PRL}, idealized mode coupling theory
\cite{Bergenholtz1999PRE}, and na\"ive  mode coupling theory \cite{Chen2004JCP}.  Simulation and theory have attributed the arrest that leads to dynamical heterogeneity to effective
particle-particle bonds arising from short-range attractive potentials \cite{Puertas2003PRE}, to
the development of entropic barrier hopping \cite{Chen2005PRE}, or to the emergence of two
distinct particle populations with long exchange times \cite{Puertas2005JPCB}.  Earlier work has
suggested a role for arrested spinodal decomposition and other equilibrium
transitions in the development of ultimate gel structure \cite{Grant1993PRE}.  Recent studies
have noted the possible perturbative effect of metastable equilibrium phase
boundaries \cite{Shah2003JPCM}, and their sensitivity to the presence of long-range charge
repulsions that are common in colloidal suspensions \cite{Groenewold2004JPCM}.  Each mechanistic
description yields particular implications for gel boundaries, structure,
kinetics, and viscoelasticity.  Thus, the relative role and interaction of
these mechanisms in model colloidal gel systems should be carefully
assessed through experiment.

Because of their availability, most quantitative comparisons to date
between theory and simulation on the one hand and experiment on the other
have focused on macroscopically measurable quantities such as gel
boundaries and collective behavior such as the dynamic structure factor.
Previously, pair dynamics of particles in gel networks have been
quantified \cite{Dinsmore2001AO}.  To further discriminate between the
competing mechanisms suggested by very recent theory and simulations \cite{Puertas2005JPCB}, 
comparison to microscopic measurements of localization length, dynamical
heterogeneity, structural populations, and barrier hopping are required.
Suggestion that rare dynamical processes play a role in gel response \cite{Chen2005PRE, Solomon2001PRE} adds another argument for measurement of
single particle distributions of dynamics.  It is the aim of this work to
fill these gaps by direct quantification of local gel structure and
dynamics through confocal microscopy.  In particular, Lu et al \cite{Lu2006PRL} and Sedgwick et al \cite{Sedgwick2004JPCM} have recently reported complex structural transitions in this regime.  The next step in this area should be to correlate structure to dynamics.   

Our gel system takes advantage of the well-characterized nature of
the polymer depletion interaction.  We choose monodisperse sterically
stabilized poly(methyl methacrylate) spheres \cite{Campbell2002JCIS} that can be
dispersed in density and refractive index matched solvent mixtures to
minimize the effects of polydispersity, sedimentation, and van der Waals
forces, which complicate experimental observation and modeling of gel
behavior.  Depletion is chosen from a range of possibilities \cite{Mohraz2005JOR, Solomon2001PRE} to provide an attractive force because it has easily tunable experimental parameters controlling both range and strength of attraction and it has been well described by theory \cite{Asakura1954JCP,
Illett1995PRE, Shah2003JCP}.  We vary the strength of the depletion
potential to determine the structural and dynamic dependence for a given
attractive range.  It has recently been reported that the range of
attraction affects the cluster morphology \cite{Lu2006PRL}.  Thus we fix this parameter at about 
four percent of the radius of the colloid in a regime where stable, tenuous gels are expected
and fine structural and dynamic changes can be differentiated.  Fluorescent particles in refractive index matched solvents lend themselves to direct observation via confocal microscopy \cite{Campbell2002JCIS, Dinsmore2001AO}, which allows local material properties to be probed.

This paper focuses on four measures that quantify both structural and
dynamical heterogeneity in gels.  For structure, we report short-range
structure by contact or bond number distributions and long-range structure
via number density fluctuations.  For dynamics we measure single particle
mean-squared displacement as well as the van Hove self-correlation
function. Here, we focus on the effects of depletion pair potential
strength on the quantities.  We find rich variation in, and correlation of,
structural and dynamic measures as the strength of pair interactions is
increased.

\section{\label{S:Methods}Methods and Materials}
\subsection{\label{particles:level2}Synthesis and characterization of monodisperse, density matched colloidal system }
The method to synthesize the poly(12-hydroxystearic acid) (PHSA) stabilized monodisperse poly(methyl methacrylate) (PMMA) colloids is as reported elsewhere \cite{Solomon2006JCP, Campbell2002JCIS, Antl1986CS}.  The fluorescent dye incorporated in the dispersion polymerization of the methyl methacrylate monomer is Nile Red ($\lambda_{ex}$= 548 nm, $\lambda_{em}$= 567 nm) \cite{Campbell2002JCIS}.  The PHSA stablizer is covalently bound to the PMMA colloids in a locking step.  After synthesis, the particles are washed 8 times in hexane, dried for at least 48 hours, characterized, and dispersed in solvents for experiments by vortex mixing and sonication. 

\begin{figure}[htbp]
\centering
\includegraphics[width=3.4in]{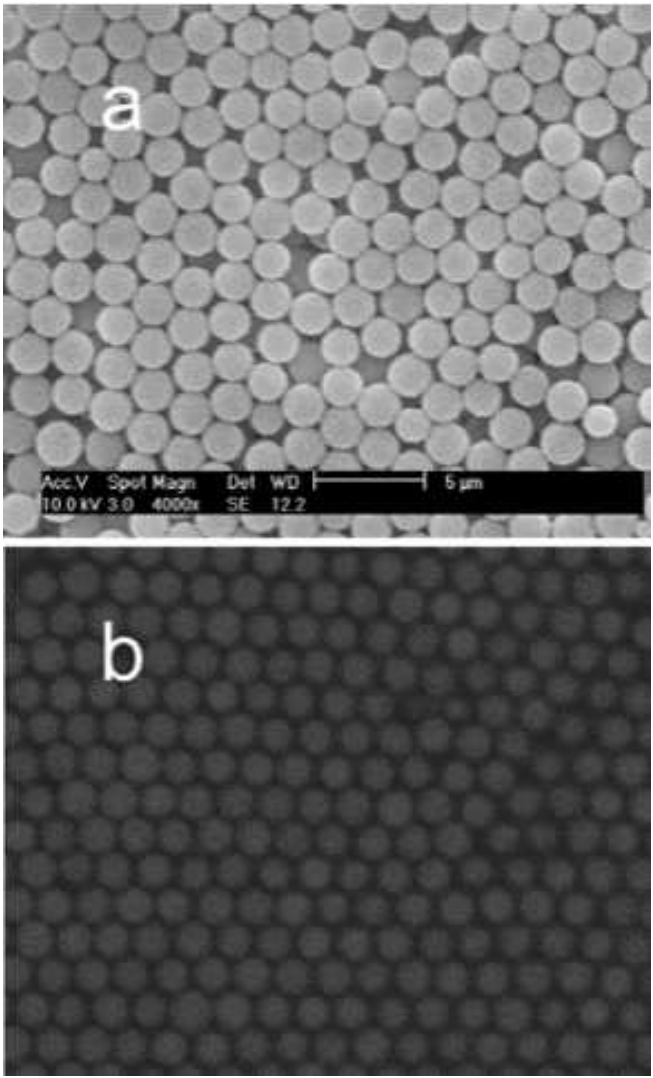}
\caption{Characterization of poly(12-hydroxysteric acid) stabilized poly(methyl methacrylate) colloids by (a) scanning electron micrographs of dry particles and (b) confocal micrographs of system crystallization at  $\phi$ = 0.49 dispersed in the solvent mixture CHB/decalin. } 
\label{f:ParticleChar}
\end{figure}

By scanning electron microscopy of the dry particles, the PMMA colloid diameter is 2.04~$\mu$m $\pm$ 3.83\% (Fig. \ref{f:ParticleChar}(a)).  By differential centrifugation, we determined a density matching solvent composition to be 65.6\% cyclohexyl bromide (CHB, 98\% purity, Sigma-Aldrich) and 34.4\% decalin (99+\% mixture of cis and trans, Sigma-Aldrich).  At these conditions, solvent equilibrated particles do not sediment after ten minutes of centrifuging at 1500 rpm (240 g).  Thus, $\frac{\Delta\rho}{\rho} \leq$ 6 $\times$ 10$^{-3}$.  The particles and solvent mixture are also nearly refractive index matched at these conditions.  The dried particles are stably redispersed in this solvent composition.  The narrow polydispersity of the colloids is demonstrated by their quiescent crystallization, as reported at $\phi$ = 0.49 in Fig. \ref{f:ParticleChar}(b).  The volume fraction range for fluid-crystal coexistence in the system is shifted relative to that for hard spheres.  This shift is consistent with the presence of charge effects \cite{Royall2003JPCM, Auer2003PRE, Yethiraj2003N}.  

\subsection{\label{charge:level2} Charge interaction measures}

The electrophoretic mobility was measured in a dilute ($\sim$1 vol \%) sample of PMMA, suspended in a density-matched mixture of cyclohexyl bromide and decalin at 25$^\circ$C using Phase Analysis Light Scattering (Malvern Instruments Zetasizer Nano, ZEN 3600).  Thirty experimental runs were performed on the sample.  The resulting value of the $\zeta$-potential was obtained by averaging the mean values of all runs.  At these solvent conditions we measured the viscosity of our solvent mixture as 2.06 cp (Schott-Ger\" ate AVS 350 viscometer) and used the value of the material dependent dielectric constant, $\epsilon_{r}$ = 5.6, as reported in literature for similar ratios of the same solvents \cite{Leunissen2005N}.  The colloid $\zeta$-potential was found to be +27 mV.

Because our system had no salt added, the ion concentration was assumed to come from the dissociation of the ions within the organic solvent mixture.  To determine the conductivity of our solvent mixture (73.9\% mass fraction CHB), we used the estimated conductivity reported previously for the no-salt case of our system \cite{Royall2005JPCM}.  At 25$^\circ$C, we obtained a conductivity of 125.7 pS/cm, by Walden's rule, implying an ion concentration of $3.0 \times 10^{-9}$ mol/L \cite{Royall2003JPCM}.  We calculate the Bjerrum length, $\lambda_{b}$ = 10.2 nm; the Debye double layer thickness, $\kappa^{-1}$ = 1.46 $\mu$m; and the surface charge, Q$_{e}$ = 165 e/colloid.  

\subsection{\label{depletant:level2} Depletion pair potential interaction due to non-adsorbing polymer}	
The non-adsoring polymer used to generate the depletion pair potential interaction is monodisperse polystyrene (Pressure Chemical, Pittsburgh, PA) of molecular weight $M_{w} = 9.0 \times 10^{5}$ g/mole and polydispersity $M_{w}/M_{n} \leq 1.10$.  At the theta condition, the radius of gyration ($R_{g}$) of this polymer would be 22 nm \cite{PolymerHandbook}.  However, we find that the density matching mixture of CHB/decalin is a good solvent for polystyrene.  This characterization was made by two independent measurements of solvent quality: mixed solvent static light scattering and intrinsic viscometry.  Characterization of $R_{g}$ and overlap concentration, c* of the non-adsorbing polymer is important because these quantities determine the range ($R_{g}$) and magnitude (c*) of the depletion interaction.  

The $R_{g}$ of polystyrene in the decalin/cyclohexyl bromide mixture was characterized by static light scattering (DAWN EOS, Wyatt Technology Corporation, Santa Barbara, CA) and application of a Zimm plot and the mixed solvent theory of Yamakawa \cite{YamakawaPolySolns}.   We determined $R_{g} = 41 \pm 4$ nm.  Applying the relationship c* = $ \frac{3M_{w}}{4\pi R_{g}^3 N_{A}}$, we compute the overlap concentration c*$_{SLS} = 0.0054$ g/mL.

The intrinsic viscosity, [$\eta$], is an indirect measurement of coil expansion and solvent quality.  Futhermore, c*[$\eta$] $\sim$ 1 \cite{LarsonComplexFluids}.  By means of capillary viscometry of 6 different polymer concentrations (Schott-Ger\" ate AVS 350 viscometer with Ubbelohde capillaries No. 52503 and 52501) we find [$\eta$] = 190 $\pm$ 0.4 mL/g at T = 22$^\circ$C.  From this value, c*$_{IV}$ =  0.0053 g/mL. The good agreement between these two independent measurements of solvent quality and coil expansion indicates that the density matched mixture of CHB/decalin is a good solvent for polystyrene.  The difference between our measured value and the theta estimate is likely related to the need to explicitly account for the mixed solvent quality effects on coil expansion as per Yamakawa \cite{YamakawaPolySolns}.  

\begin{table*}
\caption{\label{t:strength}Conditions of dimensionless polymer concentration and corresponding attractive potential between two particles in contact as calculated by the Asakura - Oosawa (AO) model of depletion for all experiments. Braces denote calculations formally outside of the range of the model.   To calculate the corresponding c/c* or U$_{AO}$/kT for this system with a theta solvent, multiply by the value 0.15.
}
\begin{ruledtabular}
\begin{tabular}{||l||c|c|c|c|c|c|c|c|c|c|c|c||}
 c/c*&0.15&0.21&0.26&0.31&0.33&0.37&0.41&0.46&0.49&0.64&1.03&1.54\\
 \hline
 U$_{AO}$/kT&7.1&9.5&11.8&14.2&15.4&17.1&19.0&21.3&22.5&29.6&\{47.4\}&\{71.1\} \\
\end{tabular}
\end{ruledtabular}
\end{table*}

In summary, based on polymer and colloid characterization, we find $\xi = R_{g}/a \approx$ 0.043 and c* $\approx$ 0.0053 g/mL.  For reference, the depletion potential strength used for experiments is listed in Table \ref{t:strength} in terms of the non-adsorbing polymer volume fraction, c/c*.  For convenience, we also report the contact pair potential of all experimental conditions, estimated using the well-known Asakura Oosawa model of depletion \cite{Asakura1954JCP, Illett1995PRE}.   More precise determination of the pair potential would be possible by application of, for example, PRISM \cite{Chatterjee1998JCP}.  

\subsection{\label{sample:level2} Sample preparation} 

Dry particles were dispersed in the density matched CHB/decalin solvent mixture by $\sim$~30~s vortex mixing and then $\sim$~10 minutes sonication.  The dispersion was quiescently equilibrated for 10 - 12 hours.  During this time, the colloid size evolves due to its interaction with the solvent.  From measurement of contact conditions of the radial distribution function, g(r), in fully gelled samples, the ultimate colloid diameter in the CHB/decalin mixture is estimated to be 1.9 $\mu$m. 

To produce suspensions with short-range attractive interactions leading to gelation, the particle dispersion was mixed for $\sim$5 s on a vortex mixer with a stock solution of non-adsorbing polymer to achieve the c/c* values of Table \ref{t:strength}.  Specimens for CLSM visualization were formed by dispensing 180 $\mu$L total volume in capillaries with an inner diameter of 6 mm.  This volume corresponds to a height to diameter ratio of 1.06.  The sample was covered by approximately 250 $\mu$L of deionized water, sealing the capillary to prevent evaporation of the solvent.

After preparation, the sample was left undisturbed on the microscope stage for a minimum of 200 minutes prior to imaging.  The kinetics of the gelation process varied between the samples, with the behavior being primarily dependent on the concentration of depletant.  Generally, in those samples where an immobilized gel structure was formed, the space-spanning network formed during the first 60 minutes after the dispersion was prepared.  Various structural rearrangements of the gel structure occurred during the following 100 minutes, during which time some of the particles and clusters within the network moved.  Once this rearrangement had occurred, the samples stabilized to a steady structural state.  We waited an additional 40 minutes before executing experiments.

\subsection{\label{CLSM:level2}  Confocal microscopy and image processing}

Gel specimens were visualized by confocal laser scanning microscopy (Leica TCS SP2) on an inverted microscope (Leica DMIRE2) using a 1.4 numerical aperture 100x oil immersion objective and an excitation wavelength of 543 nm.  At these conditions the optical resolution is $\pm$ 240 nm in the plane of the objective and $\pm$ 550 nm in the axial direction perpendicular to the objective plane.  Here the objective plane is parallel to the cover slip that forms the lower surface of the capillary.  
Structure and dynamics were quantified for all gel specimens.  For structure measurements, image volumes consisting of 247 slices at 512x512 resolution were acquired.  The pixel resolution was 146.4 nm in the x and y directions, and 122.1 nm in the z direction.  Thus, the total image volume size was: 75x75x30 $\mu m^3$.  To avoid wall effects the bottom edge of the volume was located at least 30 $\mu$m from the cover slip.  For studies of particle dynamics, time series of 500 two-dimensional 512x512 images (75 x 75 $\mu m^2$) were collected.  The time step between image acquisition was 0.848 s for a total duration of 424 s. 
Location and tracking of particles in the image volume and time series is accomplished by means of algorithms described by Crocker and Grier \cite{Crocker1996JCIS}.  The steps include a Gaussian convolution of the image pixels within a local region approximately the size of the particle radius, selection of the locally brightest pixel as a particle centroid, subpixel adjustment based on the intensity moments of the local region, and reconstruction of either a three dimensional volume for structural measures or trajectories in two dimensional time series.  

The fidelity of the particle location algorithms of the image processing was assessed by the good correspondence between colloid volume fractions computed from the added mass of colloid and from the number of particles located within an image volume for image volumes obtained in the regimes with high overlap concentrations of polymer and relatively homogenous density (average relative error = $\pm$ 4.1\%).  Furthermore, identified particle centroids were statistically uncorrelated with the location of the pixel grid.  By applying the methods of Savin and Doyle \cite{Savin2005BJ}, we determined the static error in particle location to be $\pm$ 35 nm ($\pm$ 0.24 pixel) in the objective plane and $\pm$ 45 nm ($\pm$ 0.37 slice depth) in the direction perpendicular to the objective plane.  Finally, the fidelity of the particle trajectories extracted from the image time series was assessed by comparing the self-diffusion coefficient of a 1 vol\% colloidal suspension extracted from measurement of the colloid mean-squared displacement to the predicted self-diffusivity of a free particle given the measured solvent viscosity.  This experiment, conducted in dioctyl phthalate, yielded a relative error in the self-diffusion coefficient of $\sim$ 0.1\%.  

\subsection{\label{quant:level2} Quantitative measures}

\begin{figure*}[htbp]
\centering
\includegraphics[width=5in]{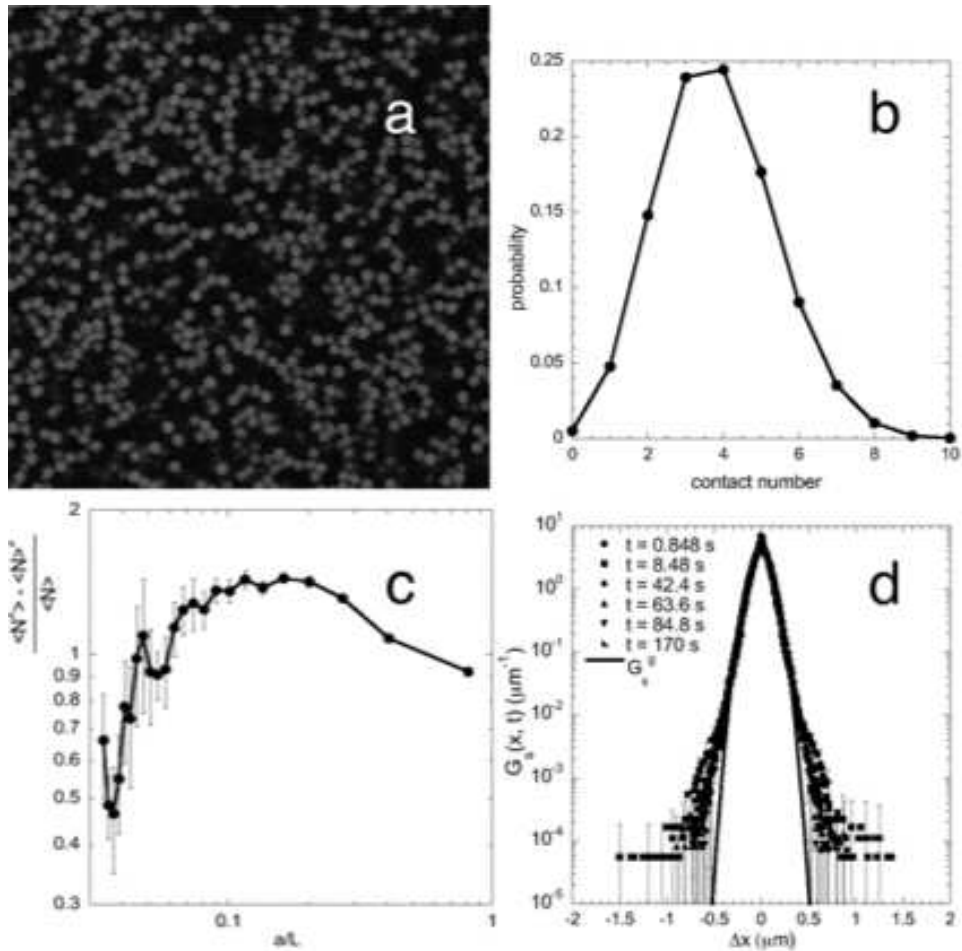}
\caption{Quantitative methods to characterize system structure and dynamics shown for  c/c* =0.64:  (a) 2D confocal image, (b) contact number probability, (c) normalized number density fluctuations, and (d) van Hove self-correlation function.  } 
\label{f:Methods}
\end{figure*}

We quantify sample structure by means of contact number distributions and number density fluctuations.  Sample dynamics are measured by means of the mean-squared displacement and van Hove self-correlation function.  These measures are briefly illustrated for a strong gel formed at c/c* = 0.64.  A sample CLSM image of this specimen is shown in Fig. \ref{f:Methods}(a). 

\subsubsection{\label{Contact:level3}Contact number distributions}
The effect of depletion-induced gelation upon the local structure is determined by analysis of contact or bond number distributions.  Two particles are considered to be in contact when the distance between their centers is less than the distance to the first minimum in the pair-correlation function, g(r) \cite{Campbell2005PRL}.  We use this method in an attractive heterogeneous environment to characterize local variation.  The first observed minimum in g(r) of the c/c* = 0.64 sample, which is considered to be a strong gel, at r/2a = 2.3 is applied to all samples for determining contact number.  

To find the mean contact number and variation for each polymer concentration, this measurement was performed on six image volumes and the average of the six mean contact numbers is reported.  Figure \ref{f:Methods}(b) shows the distribution of contact number within one sample image volume.  

\subsubsection{\label{numdens:level3}Number density fluctuations}
Within a sample volume, the long-range structure of a colloidal dispersion can be quantified by the mean-squared fluctuation of the number of particles in the system, $ \frac{<N^2>-<N>^2}{<N>}$.  Since this quantity is proportional to the isothermal compressibility of the sample, is related to the integral over all space of the pair-correlation function g(r) in the macroscopic limit, and can be related to the static structure factor in the low angle limit, S(0), it can be used to show the change in the long-range structure\cite{Varadan2003Lang}.  A maximum in the measure of number density fluctuations is indicative of a preferred cluster size.  The mean square fluctuation of the number of particles in the sample was calculated by dividing the six independent 75x75x30 $\mu m^3$ image volumes into cubic boxes of linear dimension L.  L was systematically varied from L $\sim$ a to L $\sim$ the z-dimension of the image volume.  Figure \ref{f:Methods}(c) shows sample number density fluctuation measures for c/c* = 0.64.  

\subsubsection{\label{vanHove:level3}van Hove self-correlation function and mean-squared displacement}
The self part of the van Hove correlation function, $G_{S}(x,t)$, measures the single particle displacement probability distribution \cite{HansenSimpleLiquids, Kegel2000Sci}.  In one dimension:
\begin{equation}
G_{S}(x,t) = \frac{1}{N}\left\langle
\sum_{i=1}^{N}\delta[x + x_{i}(t=0) - x_{i}(t)]\right\rangle
\label{e:vH}
\end{equation}
where N is the number of particles and $x_{i}(t)$ is particle i's position at time t.  
A free particle experiencing Brownian motion has a Gaussian form for $G_{S}(x,t)$ \cite {HansenSimpleLiquids, Kob1997PRL}:
\begin{equation}
G_{S}^{g}(x,t) = \sqrt{\frac{1}{2\pi \langle x^2 (t)\rangle}}
exp\left[\frac{-x^2}{2 \langle x^2 (t)\rangle}\right]
\label{e:vHg}
\end{equation}
where $\langle x^2 (t)\rangle$ is the one dimensional meansquared displacement.  Dynamic measurements reported in this paper are the average of 4 separate time series in different locations in the sample and error bars represent the standard error of the mean.  Figure \ref{f:Methods}(d) shows $G_{S}(x,t)$ at fixed c/c* for several instances of time.   Error bars are plotted for every fifth data point for clarity on the log scale graph.  

The one dimensional mean square displacement is the second moment of $G_{S}(x,t)$ or $\langle(x(t)-x(0))^2\rangle$ averaged over all particles in the system and is a collective measure of dynamics.  For a free particle, this value grows linearly with time as per the Stokes-Einstein equation.  Alternatively, if a particle is locally bound to a gel network, its localization length can be derived from the plateau in $\langle x^2(t)\rangle$ \cite{Chen2004JCP}.    

\section{\label{S:Results}Results}
\begin{figure*}[htbp]
\centering
\includegraphics[width=7in]{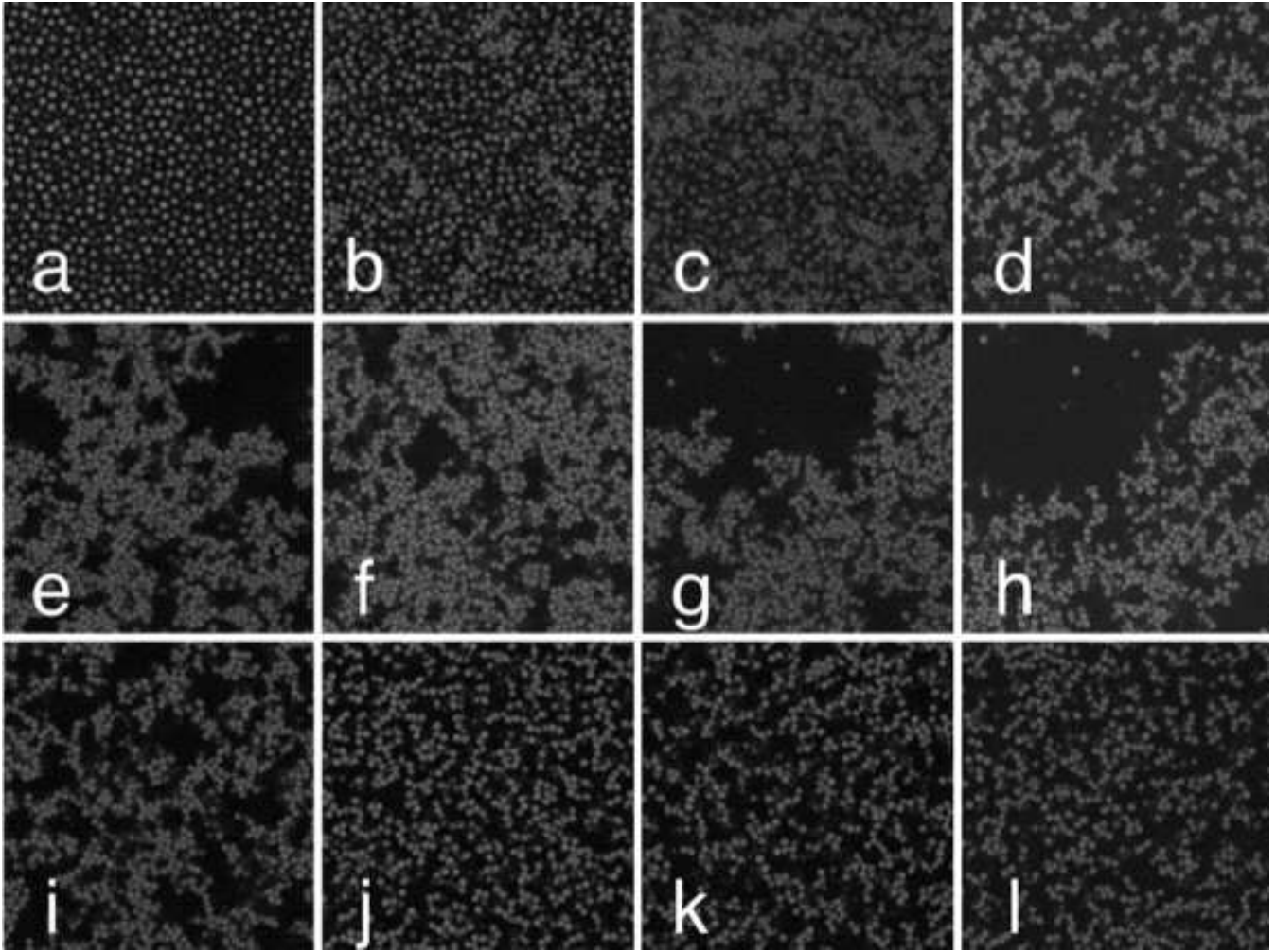}
\caption{Representative 2D confocal micrographs of system structure as a function of increasing depletion pair potential strength.  (a) c/c* = 0.15 (b) c/c* = 0.21 (c) c/c* = 0.26 (d) c/c* = 0.31 (e) c/c* = 0.33 (f) c/c* = 0.37 (g) c/c* = 0.41 (h) c/c* = 0.46 (i) c/c* = 0.49 (j) c/c* = 0.64 (k) c/c* = 1.03 (l) c/c* = 1.54.  } 
\label{f:Images}
\end{figure*}

Fig. \ref{f:Images} reports representative 2D CLSM images of structure as the magnitude of short-range interactions are increased.  At the lowest concentration shown (Fig. \ref{f:Images}(a), c/c* = 0.15), a mobile, dispersed liquid is observed.  As attraction is increased, mobile clusters form (Fig. \ref{f:Images}(b), c/c* = 0.21).  The clusters coexist with free particles.  As polymer depletant concentration is further increased, clusters become increasingly immobilized (Fig. \ref{f:Images}(c), c/c* = 0.26) until a bicontinuous network of clusters and voids is formed (Fig. \ref{f:Images}(d)-(g), c/c* = 0.31 -0.41).  At these levels of pair colloid attraction, large-scale motions in the suspensions are no longer apparent.  At c/c* $\sim$ 0.46, a region of maximum cluster/void heterogeneity is achieved.  As depletant concentration is further increased (Fig. \ref{f:Images}(i), c/c* = 0.49), structures increasingly appear more homogeneous and branched.  At the maximum concentrations studied (Fig. \ref{f:Images}(j)-(l), c/c* = 0.64 -1.53) chain like structures are apparent locally; however, globally the structure is quite homogeneous. 

\begin{figure*}[htbp]
\centering
\includegraphics[width=7in]{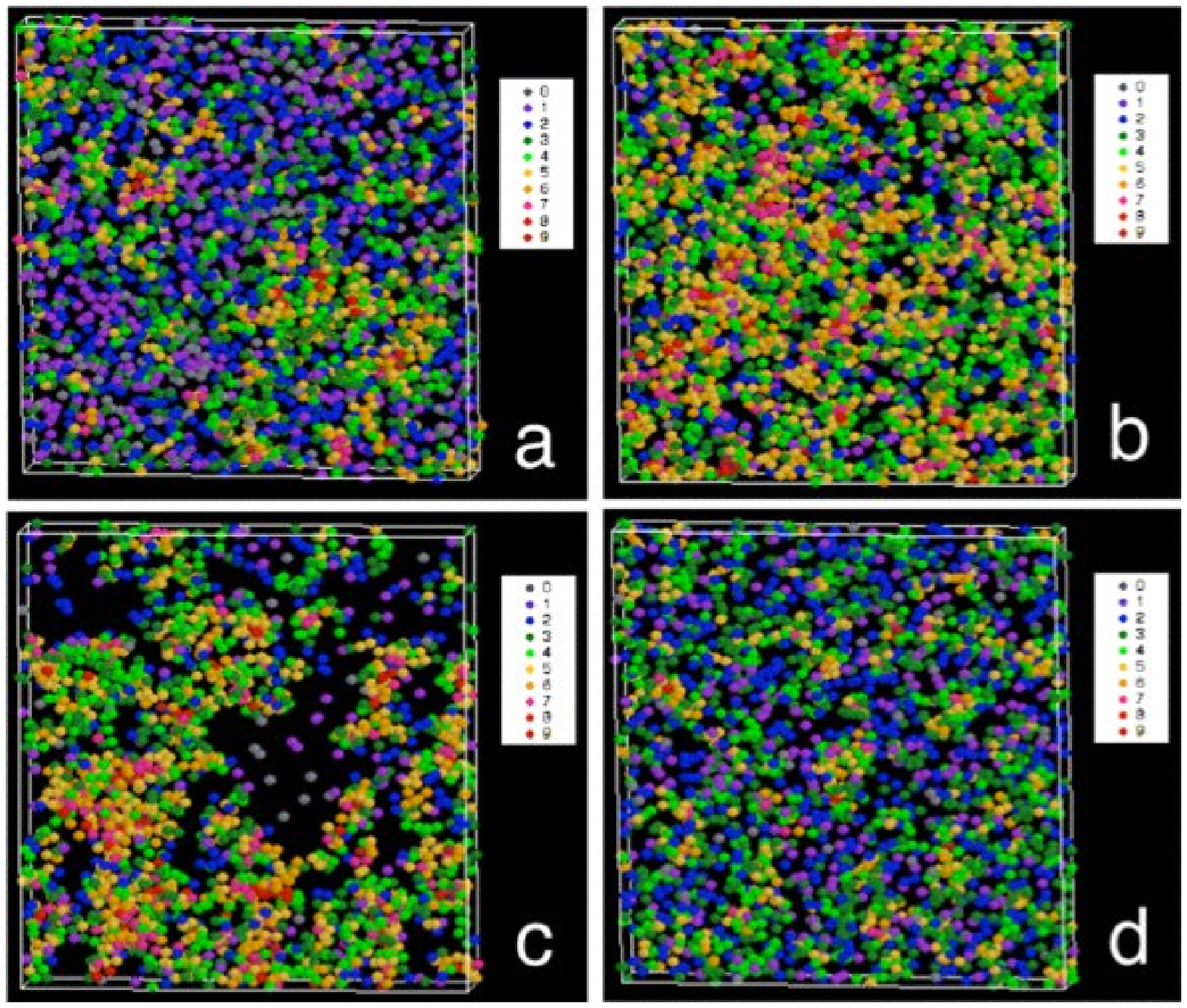}
\caption{(Color online) Color coded representation depicts spatial dependence of contact number on depletant concentration for (a) c/c* = 0.26, (b) c/c* = 0.37, (c) c/c* = 0.41, (d) c/c* = 1.03 where warmer red/orange colors correspond to high contact numbers and cooler blue/purple colors correspond to low contact numbers.  Free particles (contact number equals zero) are colored gray.  } 
\label{f:cnpics}
\end{figure*}

To quantify the local connectivity of the system, we compute contact number distributions for all c/c* studied as per the methods of Section \ref{Contact:level3}.  A color-coded representation (generated from RasMol, U-Mass) of thin ($\sim$ 5 $\mu$m) projections from the sampled image volumes is shown in Figure \ref{f:cnpics} for four different pair attraction strengths.  Fig. \ref{f:cnpics}(a) shows the system at c/c* corresponding to the immobilized cluster regime.  The cluster structure is apparent from the local regions of high contact number (orange-red coloring).  Mobile particles, with low contact probability, surround the clusters (purple-blue coloring).  The spatial extent of high contact number regions is highly variable.  In Fig. \ref{f:cnpics}(b) and (c) we see that local contact number increases in the mean.  That is, the number of low-contact number colloids progressively decreases with most being located on the surfaces of voids.  At the highest c/c*, shown in Fig. \ref{f:cnpics}(d), less spatial variability in the contact number is apparent.  Furthermore, few regions of very high contact number are found. 

\begin{figure*}[htbp]
\centering
\includegraphics[width=7in]{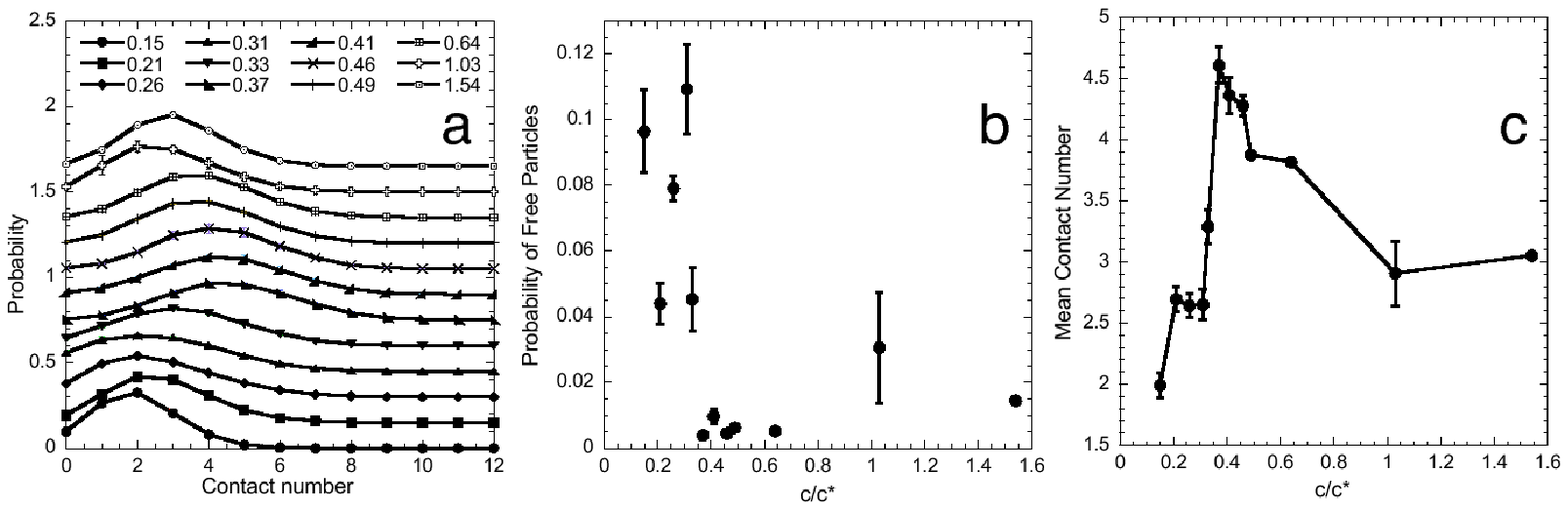}
\caption{(a) Contact number distributions for each depletant concentration.  Data are offset for clarity. (b) Probability of free particles (contact number = 0) for each sample.  (c) Mean contact number distribution for each c/c*.} 
\label{f:cnum}
\end{figure*}

These trends are apparent in the full contact number distributions plotted in Fig. \ref{f:cnum}(a) for the range of depletant concentration studied.  Data are offset for clarity.  Because of the offset, we also plot the probability of free particles (contact number = 0) for all the samples.  These data represent the points on the ordinate of Fig. \ref{f:cnum}(a).  Finally, the mean contact number of each pair attraction strength is plotted in Fig. \ref{f:cnum}(c).   The mean contact number passes through a maximum at c/c* = 0.37, where each particle is surrounded, on average, by 4.6 other particles.  The three figures quantify three qualitative observations: First, low and high pair attraction strength results in nearly the same mean contact number.  Second, intermediate pair attraction strengths (c/c* $\sim$ 0.4) yield broad contact number distributions with relatively high mean values.  Third, the number of free particles with zero contact number progressively decreases.  As many as 10\% of the colloids coexist outside aggregated clusters at c/c* $\sim$ 0.2.  By c/c* $\sim$ 0.4, the number of zero-contact number colloids detected has decreased by more than an order of magnitude.

\begin{figure*}[htbp]
\centering
\includegraphics{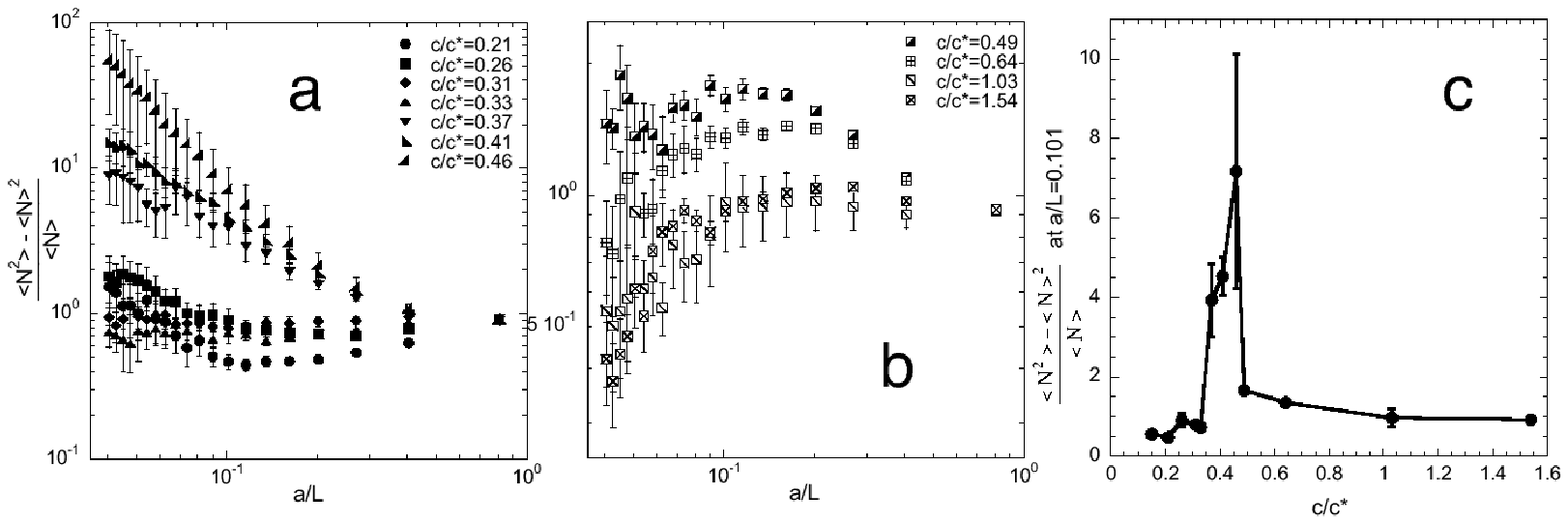}
\caption{Number density fluctuation measures show the normalized mean-squared fluctuations of particle number as a function of sub-system size considered as it varies with attraction strength for (a) c/c* $<$ 0.46 and (b) c/c* $>$ 0.49.  (c) Mean number density fluctuations for all attraction strengths for inverse length scale a/L = 0.101. } 
\label{f:numdens}
\end{figure*}

The Fig. \ref{f:Images} images show that the amount of long-range heterogeneity in the sample varies dramatically as the strength of pair attractive interactions is increased.  By varying the control parameter a/L, we assess the length scale dependence of number density fluctuations, per the method of section \ref{numdens:level3}, and therefore the degree of heterogeneity can be quantified.  Because of the non-monotonic variation of number density fluctuations with c/c*, data are plotted on two panels, Fig. \ref{f:numdens}(a) and (b).  At low c/c*, number density fluctuations are relatively small and insensitive to the inverse length scale a/L.  For comparison, the compressibility of a hard sphere suspension is $\frac{\langle N^2\rangle - \langle N\rangle^2}{\langle N\rangle}$ = 0.21 in a macroscopic limit \cite{RusselColloidalDispersions}.  For c/c* $>$ 0.35, heterogeneity rapidly increases.  The increase in compressibility is reminiscent of spinodal decomposition: if there is a characteristic length scale in the system, it has diverged beyond $\sim$ 30 colloid radii.  In this intermediate concentration range, number density fluctuations are more than two orders of magnitude greater than expected for a hard sphere fluid.  

The maximum in number density fluctuations is encountered at c/c* = 0.46.  As the strength of pair interactions is increased further, Fig. \ref{f:numdens}(b) reveals two interesting features of gel structure.  First, the magnitude of number density fluctuations collapses to values of approximately unity.  Second, a characteristic length scale, revealed as a maximum in the measure emerges at a/L $\sim$ 0.1.  This characteristic length scale is a clear indication of the emergence of a gel structure with a heterogeneous, string-like structure on intermediate length scales of about 10 colloid radii, but with a relatively homogeneous structure on longer length scales.  The full sequence of structural states is captured by Fig. \ref{f:numdens}(c), which shows how heterogeneity evolves in the system with increasing pair potential strength at the characteristic inverse length scale a/L $\sim$ 0.1.

\begin{figure}[htbp]
\centering
\includegraphics[width=3.4in]{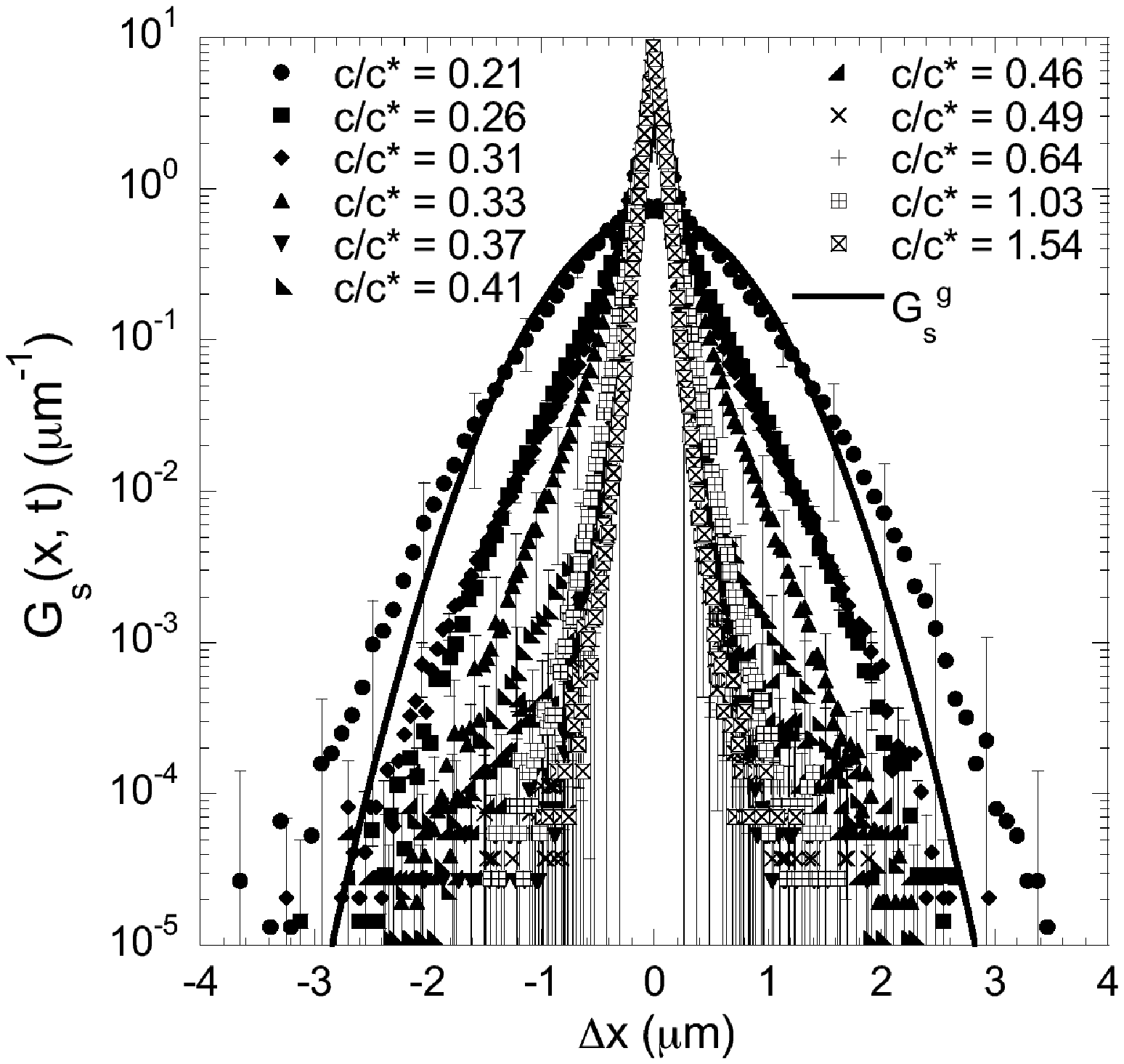}
\caption{van Hove self-correlation function for all polymer concentrations at t = 8.48 s compared to the Gaussian estimate for a free particle.  } 
\label{f:vHc}
\end{figure}

We now assess the implications of the structures reported in Figures 3-6 on system dynamics.  Using the method of Sec. \ref{vanHove:level3}, we first show the effect of c/c* on the van-Hove self-correlation function in Fig. \ref{f:vHc}.  Here data are plotted for one characteristic time (t = 8.48 s).  Data plotted for other times display the same trends.  Figure \ref{f:vHc} quantifies three trends in the distribution of single-particle displacements in the system:  First, at low c/c*, displacement distributions are nearly consistent with the Gaussian function, consistent with the observed liquid-like structure (Fig. \ref{f:Images}(a)).  Second, as c/c* is increased, single Ðparticle dynamics become increasingly localized.  The trend in c/c* is monotonic.  The greatest probability for displacement is much less than a colloid radius.  Third, as the strength of pair-attractions increased, the distribution of displacements appears increasingly non-Gaussian.

\begin{figure}[htbp]
\centering
\includegraphics[width=3.4in.]{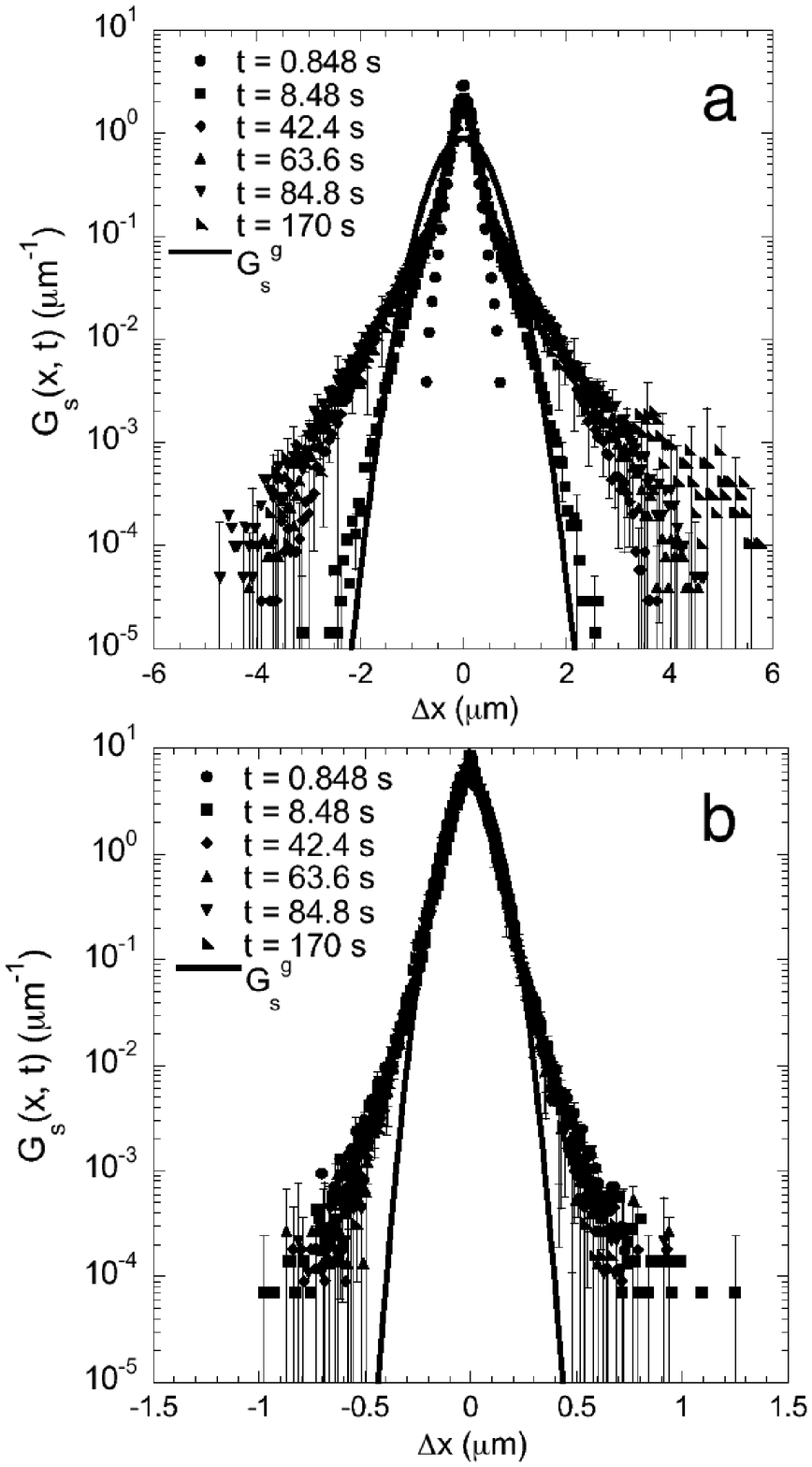}
\caption{van Hove self-correlation function at (a) low [c/c* = 0.26] and (b) high [c/c* = 1.54] depletant concentrations plotted for a range of times. } 
\label{f:vHt}
\end{figure}

Figure \ref{f:vHt} shows information complementary to Fig. \ref{f:vHc}: here at two characteristic c/c* we explore the time dependence of the distribution of single-particle displacements.  Fig. \ref{f:vHt}(a) reports data at low pair attraction strength (c/c* = 0.26), characteristic of an immobilized cluster structure.  Figure \ref{f:vHt}(b) is for a gel (c/c* = 1.54) with string-like heterogeneity.  Both samples show non-Gaussian dynamics.  At c/c* = 0.26, the Gaussian distribution appears to underpredict the distribution at both very short and very long displacements.  It overpredicts the measurements at intermediate times.  Although there is some time-dependence in the single-particle displacement distributions at c/c* = 0.26, in the strong, string-like gel regime, the dynamics are nearly time independent.

\begin{figure}[htbp]
\centering
\includegraphics[width=3.4in.]{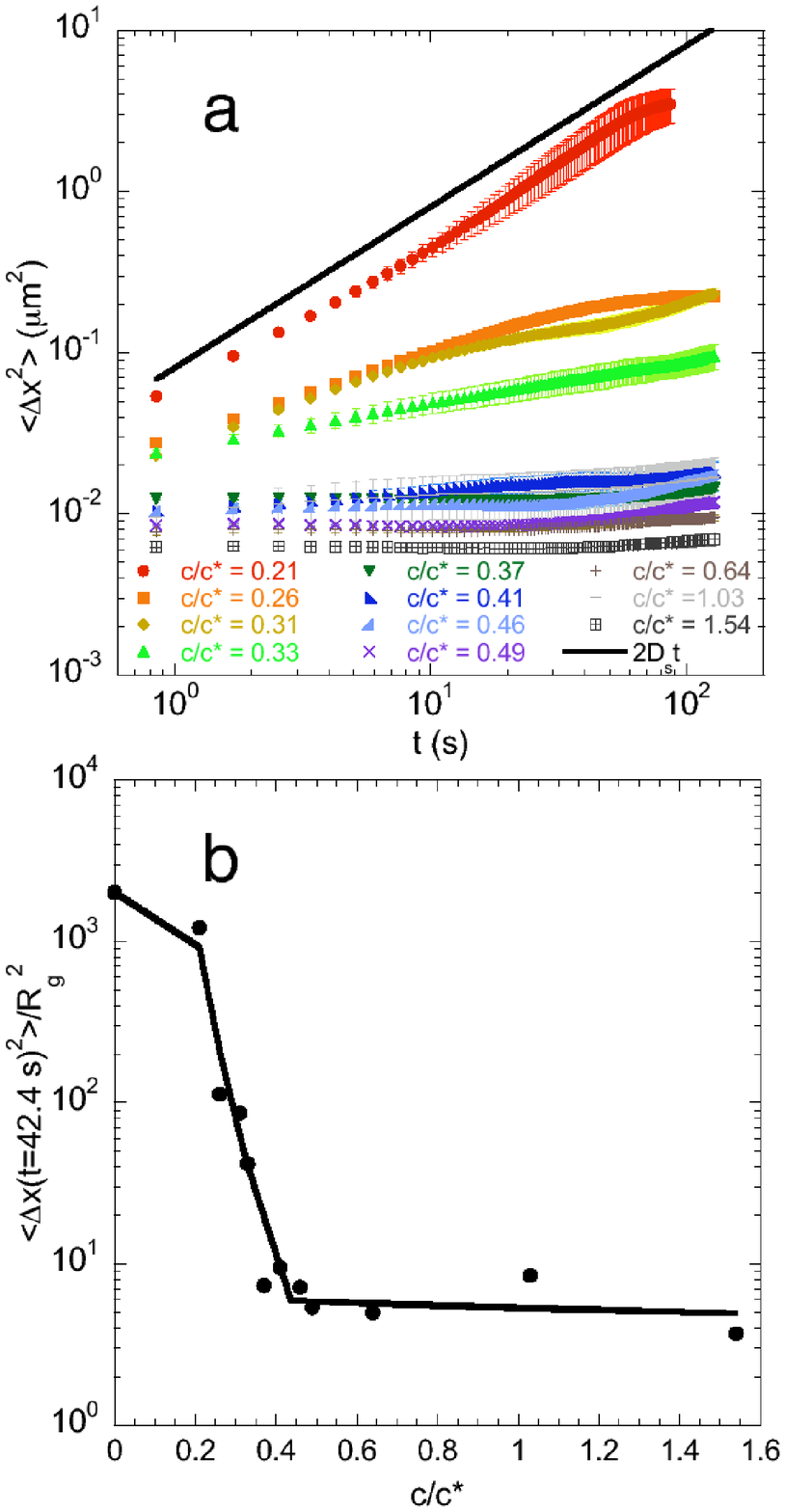}
\caption{(a) (Color online) Mean-squared displacement as a function of increasing c/c* relative to the theoretical mean-squared displacement of a particle in the equivalent medium with no non-adsorbing polymer.  (b) Localization length estimated from mean-squared displacement plateau decreases with increasing attraction point until it stabilizes at about 2.2 Rg and is no longer sensitive to concentration.  } 
\label{f:MSD}
\end{figure}

\begin{figure*}[htbp]
\centering
\includegraphics{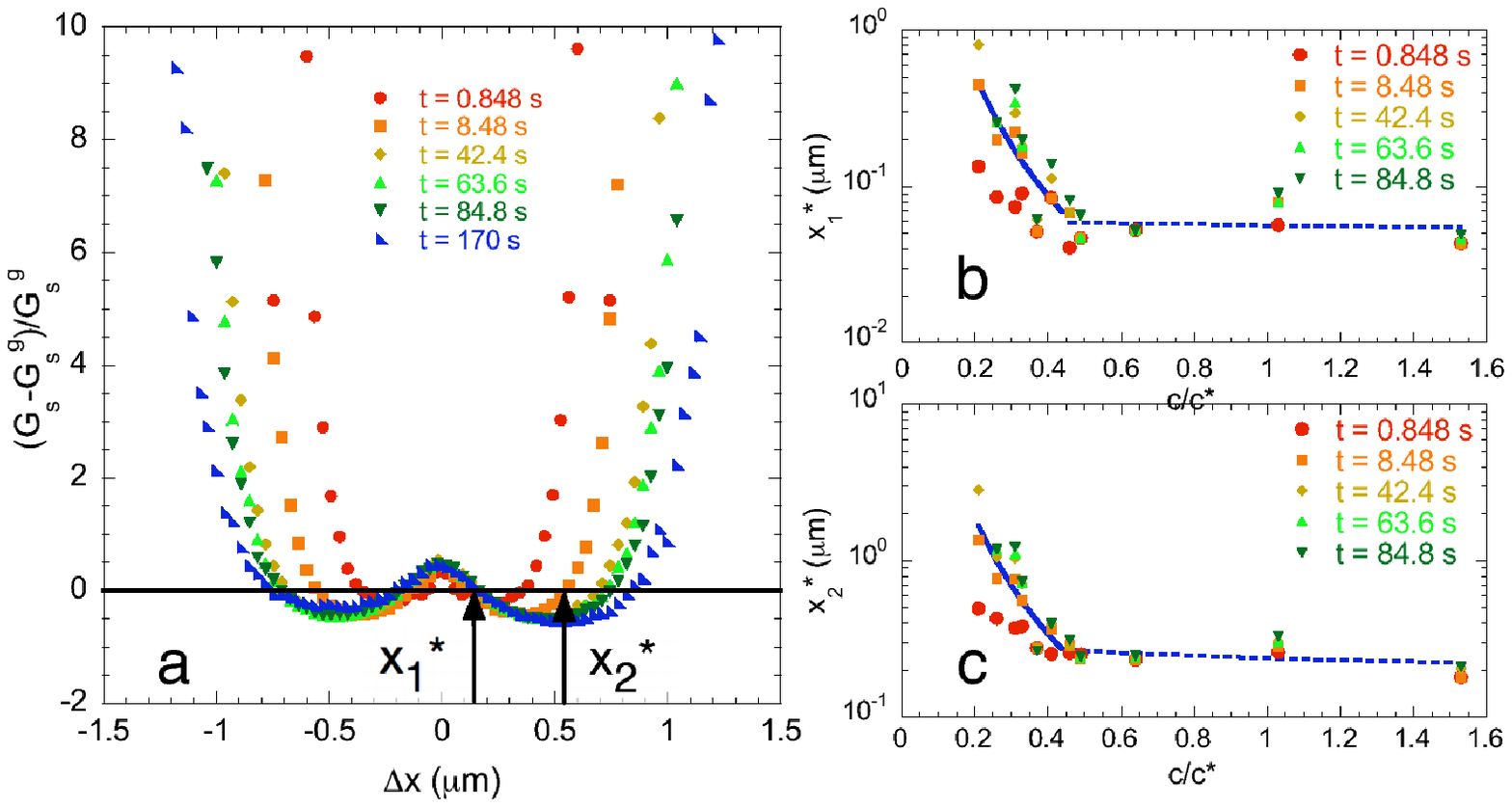}
\caption{(Color online) (a) Normalized difference between van Hove self-correlation function measured and the Gaussian estimate for the resulting mean-squared displacement for c/c* = 0.33.  x$_{1}$ represents the point where normalized difference between observed and predicted trajectories first becomes negative and x$_{2}$* represents the trajectories that move past this barrier such that there are again more trajectories observed than predicted by a Gaussian.  (b) x$_{1}$*, as a function of c/c*. (c) x$_{2}$* as a function of c/c*. } 
\label{f:Kob}
\end{figure*}

Figures \ref{f:vHc} and \ref{f:vHt} indicate the following interesting behavior that warrants further analysis: First, particles are increasingly localized as c/c* is increased.  This effect can be assessed by investigating the time-dependence of the single-particle mean-squared displacement (Fig. \ref{f:MSD}(a)).  Second, displacement distributions are highly non-Gaussian.  This effect can be assessed by investigating the deviations of $G_{S}(x,t)$ from a Gaussian function with variance equal to the measured mean-squared displacement (Fig. \ref{f:Kob}(a)).  We conclude this section by discussing these two dynamical observations.

In Fig. \ref{f:MSD}(a), the time dependence of the one-dimensional mean-squared displacement, $\langle x^2(t)\rangle$ is plotted for all the c/c* conditions.  Except for the sample with mobile cluster structure (c/c* = 0.19), the behavior at all c/c* is sub-diffusive.  Mobile clusters (c/c* = 0.19) approach the long-time self diffusivity of the colloid volume fraction $\Phi$ = 0.2 when Stokes-Einstein diffusion coefficient is modified to include deviation from the dilute limit caused by hydrodynamic interactions as per \cite{Lionberger1997JOR}.  Remaining deviation is due to the presence of clusters.  Within the temporal range probed by the experiment, the average particle mobility monotonically decreases with increasing c/c*.  The behavior is also increasingly sub-diffusive, until, under conditions that correspond to strong gels with string-like structure, the mean-squared displacement is nearly independent of time. 

Interestingly, for the strong gel structures, the localization in $\langle x^2(t)\rangle$ is roughly comparable to the square of the polymer radius of gyration, R$_{g}$, in good accord with the na\" ive/PRISM mode coupling theory prediction \cite{Chen2004JCP}.  The significance of the polymer radius of gyration here is that it sets the range of the short-range attractive pair potential between the colloids.   To fully explore this comparison further, in Fig. \ref{f:MSD}(b) we plot the plateau value of $\langle x^2(t)\rangle$ (taken at t = 42.4 s) scaled on the square of the polymer radius of gyration.  In the immobilized cluster regime (c/c* $<$ 0.26), the localization length decreases rapidly.  For stronger gels $\langle x^2(t)\rangle$ /R$_{g}^2 \sim$ 5, which corresponds to a localization length of 2.24R$_{g}$.  For comparison, Ref. \cite{Chen2004JCP} predicts r$_{localization}$/R$_{g} \sim$ 2 for a comparable system.    This first direct experimental measurement of localization length is significant because it is predicted to be a principle determinant of the linear viscoelastic response of gels \cite{Chen2004JCP}.

By empirically correlating the low c/c* and high c/c* localization behavior, as shown by the curves in Fig. \ref{f:MSD}(b), we can estimate the point of the dynamic transition to highly localized behavior.  We find the transition c/c* = 0.42.  It is instructive to compare this result to our earlier assessment of structural transitions based on the contact number (c/c* = 0.37) and number density fluctuations (c/c* = 0.46).  The agreement of these structural and dynamic transition measures suggests a key correlation between structure and dynamics in the system. 

To estimate deviations of single-particle dynamics from Gaussian behavior, we adopt the method of Kob et al \cite{Kob1997PRL} developed to analyze glassy dynamics.  The method, shown in Fig. \ref{f:Kob}(a) for a gel at c/c* = 0.33, is a quantitative way to assess dynamical heterogeneity in systems with slow dynamics.  This method clearly shows three regimes of behavior, as was qualitatively apparent in the original data of Figs. \ref{f:vHc} and \ref{f:vHt}.  At small and large displacements, positive deviations of the data from the Gaussian form are observed.  At intermediate displacement, negative deviations from the Gaussian form are observed.  Thus, we can identify two length scales that demark the regime transitions, x$_{1}$* and x$_{2}$*, as defined in Fig. \ref{f:Kob}(a).  The first, x$_{1}$*, is qualitatively similar to the localization length discussed in Fig. \ref{f:MSD}(b).  The second, x$_{2}$*, is a new scale suggested by the data.  It identifies a small population of highly mobile particles, consistent with the existence of dynamical heterogeneity in gels.  

\section{\label{S:Discussion}Discussion}

The static structure of gels formed from strong short-range interactions at low and intermediate volume fraction has recently generated intense interest.  In addition to the strength of the pair potential, structures observed are a sensitive function of potential range \cite{Lu2006PRL}, charge \cite{Sedgwick2004JPCM}, and density matching \cite{Sedgwick2004JPCM}.  Our observations of a sequence of stable fluid, mobile cluster, immobilized cluster, and gel phases as pair potential strength is increased are broadly consistent with these recent reports.  Fixed parameters in our experiments are that the colloids are charged ($\zeta$ = 165 e/colloid, $\kappa^{-1}$ = 1.46 $\mu$m), the attractive potential is short ranged ($\xi$ = 0.043), and the system is density-matched ($\Delta\rho/\rho \sim 10^{-3}$).  Our quantification of long-range number density fluctuations, short-range contact number, and single-particle dynamics allows us to build upon the phase diagrams available in Refs. \cite{Sedgwick2004JPCM} and \cite{Lu2006PRL}.

The difference between development of stable mobile and immobile cluster phases has been attributed to the degree of density matching \cite{Sedgwick2004JPCM}.  While our results do not directly test this hypothesis, the combination of structure and dynamical characterization reported here identifies the following additional features of these phases. 

First, Fig. \ref{f:MSD}(a) demonstrates that mobile and immobile cluster phases can be distinguished by the behavior of the single-particle mean-squared displacement: mobile systems (such as c/c* = 0.21) show diffusive behavior while immobile phases (0.26 $<$ c/c*$ <$ 1.54) display a plateau in the mean-squared displacement.  This quantitative metric complements previous observational discriminations and demonstrates that both types of phases can be observed at fixed density matching \cite{Sedgwick2004JPCM} and potential range \cite{Lu2006PRL}.  

Second, unlike the gel phase, where system dynamics are nearly independent of pair attraction strength (c.f. Fig. \ref{f:MSD}(b)), the immobile cluster phases dynamics are extremely sensitive to pair attractions.  For example, as the immobile cluster phase envelope is traversed, the plateau mean-squared displacement changes by about an order of magnitude. 
   
Third, the sensitivity of dynamics to pair attractions in the immobile cluster phase is correlated with significant changes in structure.  Locally, the number of free particles changes from about 10\% of the sample to about 1\% of the sample within the phase boundaries.  In addition, long-range density-fluctuations rapidly increase with pair attraction strength, in a manner reminiscent of spinodal decomposition \cite{Carpineti1992PRL}.  Note that the presence of free particles coexisting with the immobile cluster structure is consistent with the presence of charge interactions in the poly(methyl methacrylate) system \cite{Sedgwick2004JPCM}.
   
Fourth, the structural measurements indicate that the dramatic slowing down of dynamics as the immobile cluster phase envelope is traversed could conceivably be explained by a combination of two possibilities.  On the one hand, as the ratio of free to immobilized particles decreases (c.f. Fig. \ref{f:cnum}(b)) the mean system dynamics becomes increasingly skewed toward localization.  On the other hand, as pair attractions are increased, particles contained in the immobile clusters are themselves increasing bound to neighbors and thus localized \cite{Chen2004JCP}.  Analysis of the full measured van Hove self-correlation functions (c.f. Fig. \ref{f:vHc}) indicates that both mechanisms are a component of the observed changes in dynamics (data not shown).  The alternative hypothesis that the slowing down of dynamics in the immobile cluster phases is due to the quenching of cluster diffusion \cite{Kroy2004PRL} is not an explanation of the Fig. \ref{f:vHc} observation.  That is, the number-density fluctuations of Fig. \ref{f:numdens} show that cluster dimensions are larger than the acquired CLSM image volumes.  Consequently, our particle-linking algorithms, which measure motion relative to the image volume center of mass, are not sensitive to coordinated motions of this kind.   Instead, the Fig. \ref{f:vHc} results report a slowing down of the internal dynamics of the immobilized clusters.
   
 As pair attractions increase, at a critical c/c*, both measures of structure (number density fluctuation and contact number) undergo a reversal.  We identify this transition with a true gelation transition.  Structure and dynamics become increasingly insensitive to changes in pair interaction strength in the gel phase.  Interestingly, a characteristic structural length scale emerges in the system.  This scale, about 10 particle radii, is apparent from the length-scale dependence of number density fluctuations (c.f Fig. \ref{f:numdens}(b)).  The existence of long-range structural heterogeneity in colloidal gels has generated considerable controversy.  By means of scattering, Ramakrishnan et al have found that depletion gels display structural heterogeneity while thermoreversible gels do not \cite{Ramakrishnan2005Lang}.  On the other hand, Varadan and Solomon quantified long-range structural heterogeneity in a different thermoreversible gel system \cite{Varadan2003Lang}.  In the present report, measurements on well-characterized model system are instructive: the results show that, depending on the strength of pair attractions, both kinds of behavior can be found in a single system.  That is, at low pair attraction strength, long-range structural heterogeneity is apparent in the immobilized cluster phase.  At high pair attractions, structural heterogeneity is limited to intermediate scales ($\sim$ 10a).  On large scales, the gel system returns to a magnitude of number density fluctuations that is approximately consistent with a homogeneous fluid structure.  What remains for future work is to resolve how the system selects the particular length scale (in this case, $\sim$ 10a) for intermediate range structural heterogeneity.  A fundamental understanding of how this length scale develops in gels could be fruitfully applied to materials development for applications involving membranes, supports and scaffolds with particular network/void spacing.  The existence of this additional structural scale also has strong implications for both the linear \cite{Ramakrishnan2004PRE} and non-linear \cite{Mohraz2005JOR} rheological response of colloidal gels, since predicted properties vary by many orders of magnitude in response to structural heterogeneity.
 
Over a broad-range of pair attractive strengths, the depletion system displays strong departures from the Gaussian form of the van Hove self-correlation function.  This behavior is consistent with the observed long-time plateau of the single-particle mean-squared displacement for both the immobilized cluster and gel phases.  To conclude the paper, we interrogate the Fig. \ref{f:Kob} distributions to address the possibility of dynamical subpopulations in the depletion system.  In particular, Kob et al \cite{Kob1997PRL} describe a procedure to assess dynamical heterogeneity in glassy systems that can easily be applied to gel systems as well.  The method, applied to the depletion gel system, is shown in Fig. \ref{f:Kob}(a).  We see that, relative to the Gaussian reference, there is a surplus of trajectories of displacement x(t) $<$ x$_{1}$*(t), where x$_{1}$*(t) is defined as the first x where $\frac{G_{s}(x,t) - G_{S}^g(x,t)} {G_{S}^g(x,t)}$ becomes negative.  This observation is consistent with localization.  That is, an excess of trajectories are limited to x $<$ x$_{1}$*(t).  Indeed, in the gel phase, x$_{1}$*(t) is approximately the range of attractive depletion potential.  Localization on this scale is predicted by mode coupling theories of gelation \cite{Bergenholtz2003Lang, Chen2004JCP}.
 
If localization were the only consequence of the slowing down of dynamics in depletion gelation, then x$_{1}$*(t) would be the only dynamical length scale.  However, Fig. \ref{f:Kob} suggests an additional dynamical length scale x$_{2}$*(t).  That is, in both the immobilized cluster and gel phases, the subpopulation of colloids with displacements x(t) $>$ x$_{2}$*(t) is anomalously large relative to the corresponding Gaussian distribution.  This observation suggests that an additional mechanism besides the localization predicted by mode coupling theory underlies the systemÕs dynamics.
 
This region at large displacements of positive deviations from Gaussian behavior typically comprises about 1-3\% of the systems' total trajectories.  In glasses, behavior of this kind is an indicator of dynamical heterogeneity, the non-exponential decay of time correlation functions due to the superposition of individual exponential relaxations with different time constants.  Dynamical heterogeneity has been extensively studied in attractive systems by means of simulation \cite{Puertas2003PRE}; however, to our knowledge Fig. \ref{f:Kob} is the first experimental report of its presence in attractive systems.  Thus, the results represent a useful confirmation that the complex population dynamics of gels observed in simulations is indeed consistent with the behavior of real systems. 

To conclude, we examine hypothetical origins of the observed dynamical heterogeneity.  First, in the immobilized cluster phase, one contribution might be the simple superposition of the slow internal dynamics of the clusters with the fast diffusion of free particles co-existing with the clusters.  However this explanation is not applicable to the gel phase because it lacks a significant population of free particles.  Alternatively, activated hopping transitions are a driver of rare, large displacement dynamics in glassy systems and could hypothetically arise in the attractive system studied here \cite{Chen2005PRE}.  In addition, Puertas et al \cite{Puertas2005JPCB} discuss the possibility of a collective structure origin to dynamical heterogeneity.  That is, the emergence of structural heterogeneity, such as associated with incipient spinodal decomposition or equilibrium clustering \cite{Kumar2001PRL, Zaccarelli2005PRL}, could be correlated with dynamical heterogeneity.  With regards to this last hypothesis, we note the coincidence of the system's collective structure transition (Fig. \ref{f:numdens}(c)) and the transition in the dynamical heterogeneity length scale x$_{2}$* (Fig. \ref{f:Kob}(c)).  In addition, the strength of pair attractions appears to affect collective structure in the immobile cluster and gel phases in a way that is similar to its effect on dynamical heterogeneity.   That is, collective structure and dynamical heterogeneity are much more sensitive to the strength of pair attractions in the immobilized cluster phase than in the gel phase.  Thus, although we cannot at this time definitively resolve among the listed hypotheses, we note that the comparison of the features of Fig. \ref{f:numdens}(c) and Fig. \ref{f:Kob}(c) are highly suggestive of a role for collective structure in mediating dynamical heterogeneity in attractive gel systems at intermediate volume fractions.
 
\section*{\label{S:Ack}Acknowledgments}  

We acknowledge Hagar Zohar for assistance with viscosity measurements and Timothy Lewer for assistance with particle sizing.  We appreciate useful discussions on particle tracking with Eric Weeks, Gianguido Cianci, and Thierry Savin, and production of glass capillaries by Harald Eberhart.  The National Science Foundation (CTS-0093076 and CTS-0522340) supported this work.  

\bibliography{struct_dyn}

\begin{thebibliography}{60}
\expandafter\ifx\csname natexlab\endcsname\relax\def\natexlab#1{#1}\fi
\expandafter\ifx\csname bibnamefont\endcsname\relax
  \def\bibnamefont#1{#1}\fi
\expandafter\ifx\csname bibfnamefont\endcsname\relax
  \def\bibfnamefont#1{#1}\fi
\expandafter\ifx\csname citenamefont\endcsname\relax
  \def\citenamefont#1{#1}\fi
\expandafter\ifx\csname url\endcsname\relax
  \def\url#1{\texttt{#1}}\fi
\expandafter\ifx\csname urlprefix\endcsname\relax\def\urlprefix{URL }\fi
\providecommand{\bibinfo}[2]{#2}
\providecommand{\eprint}[2][]{\url{#2}}

\bibitem[{\citenamefont{Mezzenga et~al.}(2005)\citenamefont{Mezzenga,
  Schurtenberger, Burbidge, and Michel}}]{SoftFoodReview2005NatMat}
\bibinfo{author}{\bibfnamefont{R.}~\bibnamefont{Mezzenga}},
  \bibinfo{author}{\bibfnamefont{P.}~\bibnamefont{Schurtenberger}},
  \bibinfo{author}{\bibfnamefont{A.}~\bibnamefont{Burbidge}}, \bibnamefont{and}
  \bibinfo{author}{\bibfnamefont{M.}~\bibnamefont{Michel}},
  \bibinfo{journal}{Nature Materials} \textbf{\bibinfo{volume}{4}},
  \bibinfo{pages}{729} (\bibinfo{year}{2005}).

\bibitem[{\citenamefont{Shanbhag et~al.}(2005)\citenamefont{Shanbhag, Wang, and
  Kotov}}]{Shanbhag2005Small}
\bibinfo{author}{\bibfnamefont{S.}~\bibnamefont{Shanbhag}},
  \bibinfo{author}{\bibfnamefont{S.~P.} \bibnamefont{Wang}}, \bibnamefont{and}
  \bibinfo{author}{\bibfnamefont{N.~A.} \bibnamefont{Kotov}},
  \bibinfo{journal}{Small} \textbf{\bibinfo{volume}{1}}, \bibinfo{pages}{1208}
  (\bibinfo{year}{2005}).

\bibitem[{\citenamefont{Conrad et~al.}(2005)\citenamefont{Conrad, Starr, and
  Weitz}}]{Conrad2005JPCB}
\bibinfo{author}{\bibfnamefont{J.~C.} \bibnamefont{Conrad}},
  \bibinfo{author}{\bibfnamefont{F.~W.} \bibnamefont{Starr}}, \bibnamefont{and}
  \bibinfo{author}{\bibfnamefont{D.~A.} \bibnamefont{Weitz}},
  \bibinfo{journal}{Journal of Physical Chemistry B}
  \textbf{\bibinfo{volume}{109}}, \bibinfo{pages}{21235}
  (\bibinfo{year}{2005}).

\bibitem[{\citenamefont{Kegel and van Blaaderen}(2000)}]{Kegel2000Sci}
\bibinfo{author}{\bibfnamefont{W.~K.} \bibnamefont{Kegel}} \bibnamefont{and}
  \bibinfo{author}{\bibfnamefont{A.}~\bibnamefont{van Blaaderen}},
  \bibinfo{journal}{Science} \textbf{\bibinfo{volume}{287}},
  \bibinfo{pages}{290} (\bibinfo{year}{2000}).

\bibitem[{\citenamefont{Weeks et~al.}(2000)\citenamefont{Weeks, Crocker,
  Levitt, Schofield, and Weitz}}]{Weeks2000Sci}
\bibinfo{author}{\bibfnamefont{E.~R.} \bibnamefont{Weeks}},
  \bibinfo{author}{\bibfnamefont{J.~C.} \bibnamefont{Crocker}},
  \bibinfo{author}{\bibfnamefont{A.~C.} \bibnamefont{Levitt}},
  \bibinfo{author}{\bibfnamefont{A.}~\bibnamefont{Schofield}},
  \bibnamefont{and} \bibinfo{author}{\bibfnamefont{D.~A.} \bibnamefont{Weitz}},
  \bibinfo{journal}{Science} \textbf{\bibinfo{volume}{287}},
  \bibinfo{pages}{627} (\bibinfo{year}{2000}).

\bibitem[{\citenamefont{Segre et~al.}(2001)\citenamefont{Segre, Prasad,
  Schofield, and Weitz}}]{Segre2001PRL}
\bibinfo{author}{\bibfnamefont{P.~N.} \bibnamefont{Segre}},
  \bibinfo{author}{\bibfnamefont{V.}~\bibnamefont{Prasad}},
  \bibinfo{author}{\bibfnamefont{A.~B.} \bibnamefont{Schofield}},
  \bibnamefont{and} \bibinfo{author}{\bibfnamefont{D.~A.} \bibnamefont{Weitz}},
  \bibinfo{journal}{Physical Review Letters} \textbf{\bibinfo{volume}{86}},
  \bibinfo{pages}{6042} (\bibinfo{year}{2001}).

\bibitem[{\citenamefont{Carpineti and Giglio}(1992)}]{Carpineti1992PRL}
\bibinfo{author}{\bibfnamefont{M.}~\bibnamefont{Carpineti}} \bibnamefont{and}
  \bibinfo{author}{\bibfnamefont{M.}~\bibnamefont{Giglio}},
  \bibinfo{journal}{Physical Review Letters} \textbf{\bibinfo{volume}{68}},
  \bibinfo{pages}{3327} (\bibinfo{year}{1992}).

\bibitem[{\citenamefont{Pham et~al.}(2002)\citenamefont{Pham, Puertas,
  Bergenholtz, Egelhaaf, Moussaid, Pusey, Schofield, Cates, Fuchs, and
  Poon}}]{Pham2002Sci}
\bibinfo{author}{\bibfnamefont{K.~N.} \bibnamefont{Pham}},
  \bibinfo{author}{\bibfnamefont{A.~M.} \bibnamefont{Puertas}},
  \bibinfo{author}{\bibfnamefont{J.}~\bibnamefont{Bergenholtz}},
  \bibinfo{author}{\bibfnamefont{S.~U.} \bibnamefont{Egelhaaf}},
  \bibinfo{author}{\bibfnamefont{A.}~\bibnamefont{Moussaid}},
  \bibinfo{author}{\bibfnamefont{P.~N.} \bibnamefont{Pusey}},
  \bibinfo{author}{\bibfnamefont{A.~B.} \bibnamefont{Schofield}},
  \bibinfo{author}{\bibfnamefont{M.~E.} \bibnamefont{Cates}},
  \bibinfo{author}{\bibfnamefont{M.}~\bibnamefont{Fuchs}}, \bibnamefont{and}
  \bibinfo{author}{\bibfnamefont{W.~C.~K.} \bibnamefont{Poon}},
  \bibinfo{journal}{Science} \textbf{\bibinfo{volume}{296}},
  \bibinfo{pages}{104} (\bibinfo{year}{2002}).

\bibitem[{\citenamefont{Weitz et~al.}(1984)\citenamefont{Weitz, Huang, Lin, and
  Sung}}]{Weitz1984PRL}
\bibinfo{author}{\bibfnamefont{D.~A.} \bibnamefont{Weitz}},
  \bibinfo{author}{\bibfnamefont{J.~S.} \bibnamefont{Huang}},
  \bibinfo{author}{\bibfnamefont{M.~Y.} \bibnamefont{Lin}}, \bibnamefont{and}
  \bibinfo{author}{\bibfnamefont{J.}~\bibnamefont{Sung}},
  \bibinfo{journal}{Physical Review Letters} \textbf{\bibinfo{volume}{53}},
  \bibinfo{pages}{1657} (\bibinfo{year}{1984}).

\bibitem[{\citenamefont{Lu et~al.}(2006)\citenamefont{Lu, Conrad, Wyss,
  Schofield, and Weitz}}]{Lu2006PRL}
\bibinfo{author}{\bibfnamefont{P.~J.} \bibnamefont{Lu}},
  \bibinfo{author}{\bibfnamefont{J.~C.} \bibnamefont{Conrad}},
  \bibinfo{author}{\bibfnamefont{H.~M.} \bibnamefont{Wyss}},
  \bibinfo{author}{\bibfnamefont{A.~B.} \bibnamefont{Schofield}},
  \bibnamefont{and} \bibinfo{author}{\bibfnamefont{D.~A.} \bibnamefont{Weitz}},
  \bibinfo{journal}{Physical Review Letters} \textbf{\bibinfo{volume}{96}},
  \bibinfo{pages}{028306} (\bibinfo{year}{2006}).

\bibitem[{\citenamefont{Ramakrishnan et~al.}(2005)\citenamefont{Ramakrishnan,
  Gopalakrishnan, and Zukoski}}]{Ramakrishnan2005Lang}
\bibinfo{author}{\bibfnamefont{S.}~\bibnamefont{Ramakrishnan}},
  \bibinfo{author}{\bibfnamefont{V.}~\bibnamefont{Gopalakrishnan}},
  \bibnamefont{and} \bibinfo{author}{\bibfnamefont{C.~F.}
  \bibnamefont{Zukoski}}, \bibinfo{journal}{Langmuir}
  \textbf{\bibinfo{volume}{21}}, \bibinfo{pages}{9917} (\bibinfo{year}{2005}).

\bibitem[{\citenamefont{Sedgwick et~al.}(2004)\citenamefont{Sedgwick, Egelhaaf,
  and Poon}}]{Sedgwick2004JPCM}
\bibinfo{author}{\bibfnamefont{H.}~\bibnamefont{Sedgwick}},
  \bibinfo{author}{\bibfnamefont{S.~U.} \bibnamefont{Egelhaaf}},
  \bibnamefont{and} \bibinfo{author}{\bibfnamefont{W.~C.~K.}
  \bibnamefont{Poon}}, \bibinfo{journal}{Journal of Physics-Condensed Matter}
  \textbf{\bibinfo{volume}{16}}, \bibinfo{pages}{S4913} (\bibinfo{year}{2004}).

\bibitem[{\citenamefont{Stradner et~al.}(2004)\citenamefont{Stradner, Sedgwick,
  Cardinaux, Poon, Egelhaaf, and Schurtenberger}}]{Stradner2004Nat}
\bibinfo{author}{\bibfnamefont{A.}~\bibnamefont{Stradner}},
  \bibinfo{author}{\bibfnamefont{H.}~\bibnamefont{Sedgwick}},
  \bibinfo{author}{\bibfnamefont{F.}~\bibnamefont{Cardinaux}},
  \bibinfo{author}{\bibfnamefont{W.~C.~K.} \bibnamefont{Poon}},
  \bibinfo{author}{\bibfnamefont{S.~U.} \bibnamefont{Egelhaaf}},
  \bibnamefont{and}
  \bibinfo{author}{\bibfnamefont{P.}~\bibnamefont{Schurtenberger}},
  \bibinfo{journal}{Nature} \textbf{\bibinfo{volume}{432}},
  \bibinfo{pages}{492} (\bibinfo{year}{2004}).

\bibitem[{\citenamefont{Shah et~al.}(2003{\natexlab{a}})\citenamefont{Shah,
  Chen, Ramakrishnan, Schweizer, and Zukoski}}]{Shah2003JPCM}
\bibinfo{author}{\bibfnamefont{S.~A.} \bibnamefont{Shah}},
  \bibinfo{author}{\bibfnamefont{Y.~L.} \bibnamefont{Chen}},
  \bibinfo{author}{\bibfnamefont{S.}~\bibnamefont{Ramakrishnan}},
  \bibinfo{author}{\bibfnamefont{K.~S.} \bibnamefont{Schweizer}},
  \bibnamefont{and} \bibinfo{author}{\bibfnamefont{C.~F.}
  \bibnamefont{Zukoski}}, \bibinfo{journal}{Journal of Physics-Condensed
  Matter} \textbf{\bibinfo{volume}{15}}, \bibinfo{pages}{4751}
  (\bibinfo{year}{2003}{\natexlab{a}}).

\bibitem[{\citenamefont{Varadan and Solomon}(2003)}]{Varadan2003Lang}
\bibinfo{author}{\bibfnamefont{P.}~\bibnamefont{Varadan}} \bibnamefont{and}
  \bibinfo{author}{\bibfnamefont{M.~J.} \bibnamefont{Solomon}},
  \bibinfo{journal}{Langmuir} \textbf{\bibinfo{volume}{19}},
  \bibinfo{pages}{509} (\bibinfo{year}{2003}).

\bibitem[{\citenamefont{Krall and Weitz}(1998)}]{Krall1998PRL}
\bibinfo{author}{\bibfnamefont{A.~H.} \bibnamefont{Krall}} \bibnamefont{and}
  \bibinfo{author}{\bibfnamefont{D.~A.} \bibnamefont{Weitz}},
  \bibinfo{journal}{Physical Review Letters} \textbf{\bibinfo{volume}{80}},
  \bibinfo{pages}{778} (\bibinfo{year}{1998}).

\bibitem[{\citenamefont{Cates et~al.}(2004)\citenamefont{Cates, Fuchs, Kroy,
  Poon, and Puertas}}]{Cates2004JPCM}
\bibinfo{author}{\bibfnamefont{M.~E.} \bibnamefont{Cates}},
  \bibinfo{author}{\bibfnamefont{M.}~\bibnamefont{Fuchs}},
  \bibinfo{author}{\bibfnamefont{K.}~\bibnamefont{Kroy}},
  \bibinfo{author}{\bibfnamefont{W.~C.~K.} \bibnamefont{Poon}},
  \bibnamefont{and} \bibinfo{author}{\bibfnamefont{A.~M.}
  \bibnamefont{Puertas}}, \bibinfo{journal}{Journal of Physics-Condensed
  Matter} \textbf{\bibinfo{volume}{16}}, \bibinfo{pages}{S4861}
  (\bibinfo{year}{2004}).

\bibitem[{\citenamefont{Manley et~al.}(2005)\citenamefont{Manley, Wyss,
  Miyazaki, Conrad, Trappe, Kaufman, Reichman, and Weitz}}]{Manley2005PRL}
\bibinfo{author}{\bibfnamefont{S.}~\bibnamefont{Manley}},
  \bibinfo{author}{\bibfnamefont{H.~M.} \bibnamefont{Wyss}},
  \bibinfo{author}{\bibfnamefont{K.}~\bibnamefont{Miyazaki}},
  \bibinfo{author}{\bibfnamefont{J.~C.} \bibnamefont{Conrad}},
  \bibinfo{author}{\bibfnamefont{V.}~\bibnamefont{Trappe}},
  \bibinfo{author}{\bibfnamefont{L.~J.} \bibnamefont{Kaufman}},
  \bibinfo{author}{\bibfnamefont{D.~R.} \bibnamefont{Reichman}},
  \bibnamefont{and} \bibinfo{author}{\bibfnamefont{D.~A.} \bibnamefont{Weitz}},
  \bibinfo{journal}{Physical Review Letters} \textbf{\bibinfo{volume}{95}},
  \bibinfo{pages}{238302} (\bibinfo{year}{2005}).

\bibitem[{\citenamefont{Verduin and Dhont}(1995)}]{Verduin1995JCIS}
\bibinfo{author}{\bibfnamefont{H.}~\bibnamefont{Verduin}} \bibnamefont{and}
  \bibinfo{author}{\bibfnamefont{J.~K.~G.} \bibnamefont{Dhont}},
  \bibinfo{journal}{Journal of Colloid and Interface Science}
  \textbf{\bibinfo{volume}{172}}, \bibinfo{pages}{425} (\bibinfo{year}{1995}).

\bibitem[{\citenamefont{Grant and Russel}(1993)}]{Grant1993PRE}
\bibinfo{author}{\bibfnamefont{M.~C.} \bibnamefont{Grant}} \bibnamefont{and}
  \bibinfo{author}{\bibfnamefont{W.~B.} \bibnamefont{Russel}},
  \bibinfo{journal}{Physical Review E} \textbf{\bibinfo{volume}{47}},
  \bibinfo{pages}{2606} (\bibinfo{year}{1993}).

\bibitem[{\citenamefont{Elliott et~al.}(2003)\citenamefont{Elliott, Butera,
  Hanus, and Wagner}}]{Elliott2003FarD}
\bibinfo{author}{\bibfnamefont{S.~L.} \bibnamefont{Elliott}},
  \bibinfo{author}{\bibfnamefont{R.~J.} \bibnamefont{Butera}},
  \bibinfo{author}{\bibfnamefont{L.~H.} \bibnamefont{Hanus}}, \bibnamefont{and}
  \bibinfo{author}{\bibfnamefont{N.~J.} \bibnamefont{Wagner}},
  \bibinfo{journal}{Faraday Disscussions} \textbf{\bibinfo{volume}{123}},
  \bibinfo{pages}{369} (\bibinfo{year}{2003}).

\bibitem[{\citenamefont{Campbell et~al.}(2005)\citenamefont{Campbell, Anderson,
  van Duijneveldt, and Bartlett}}]{Campbell2005PRL}
\bibinfo{author}{\bibfnamefont{A.~I.} \bibnamefont{Campbell}},
  \bibinfo{author}{\bibfnamefont{V.~J.} \bibnamefont{Anderson}},
  \bibinfo{author}{\bibfnamefont{J.~S.} \bibnamefont{van Duijneveldt}},
  \bibnamefont{and} \bibinfo{author}{\bibfnamefont{P.}~\bibnamefont{Bartlett}},
  \bibinfo{journal}{Physical Review Letters} \textbf{\bibinfo{volume}{94}},
  \bibinfo{pages}{208301} (\bibinfo{year}{2005}).

\bibitem[{\citenamefont{Pham et~al.}(2004)\citenamefont{Pham, Egelhaaf, Pusey,
  and Poon}}]{Pham2004PRE}
\bibinfo{author}{\bibfnamefont{K.~N.} \bibnamefont{Pham}},
  \bibinfo{author}{\bibfnamefont{S.~U.} \bibnamefont{Egelhaaf}},
  \bibinfo{author}{\bibfnamefont{P.~N.} \bibnamefont{Pusey}}, \bibnamefont{and}
  \bibinfo{author}{\bibfnamefont{W.~C.~K.} \bibnamefont{Poon}},
  \bibinfo{journal}{Physical Review E} \textbf{\bibinfo{volume}{69}},
  \bibinfo{pages}{011503} (\bibinfo{year}{2004}).

\bibitem[{\citenamefont{Pusey et~al.}(1993)\citenamefont{Pusey, Pirie, and
  Poon}}]{Pusey1993Pa}
\bibinfo{author}{\bibfnamefont{P.~N.} \bibnamefont{Pusey}},
  \bibinfo{author}{\bibfnamefont{A.~D.} \bibnamefont{Pirie}}, \bibnamefont{and}
  \bibinfo{author}{\bibfnamefont{W.~C.~K.} \bibnamefont{Poon}},
  \bibinfo{journal}{Physica A:Statistical Mechanics and Its Applications}
  \textbf{\bibinfo{volume}{201}}, \bibinfo{pages}{322} (\bibinfo{year}{1993}).

\bibitem[{\citenamefont{Puertas et~al.}(2002)\citenamefont{Puertas, Fuchs, and
  Cates}}]{Puertas2002PRL}
\bibinfo{author}{\bibfnamefont{A.~M.} \bibnamefont{Puertas}},
  \bibinfo{author}{\bibfnamefont{M.}~\bibnamefont{Fuchs}}, \bibnamefont{and}
  \bibinfo{author}{\bibfnamefont{M.~E.} \bibnamefont{Cates}},
  \bibinfo{journal}{Physical Review Letters} \textbf{\bibinfo{volume}{88}},
  \bibinfo{pages}{098301} (\bibinfo{year}{2002}).

\bibitem[{\citenamefont{Bergenholtz and Fuchs}(1999)}]{Bergenholtz1999PRE}
\bibinfo{author}{\bibfnamefont{J.}~\bibnamefont{Bergenholtz}} \bibnamefont{and}
  \bibinfo{author}{\bibfnamefont{M.}~\bibnamefont{Fuchs}},
  \bibinfo{journal}{Physical Review E} \textbf{\bibinfo{volume}{59}},
  \bibinfo{pages}{5706} (\bibinfo{year}{1999}).

\bibitem[{\citenamefont{Chen and Schweizer}(2004)}]{Chen2004JCP}
\bibinfo{author}{\bibfnamefont{Y.~L.} \bibnamefont{Chen}} \bibnamefont{and}
  \bibinfo{author}{\bibfnamefont{K.~S.} \bibnamefont{Schweizer}},
  \bibinfo{journal}{Journal of Chemical Physics}
  \textbf{\bibinfo{volume}{120}}, \bibinfo{pages}{7212} (\bibinfo{year}{2004}).

\bibitem[{\citenamefont{Puertas et~al.}(2003)\citenamefont{Puertas, Fuchs, and
  Cates}}]{Puertas2003PRE}
\bibinfo{author}{\bibfnamefont{A.~M.} \bibnamefont{Puertas}},
  \bibinfo{author}{\bibfnamefont{M.}~\bibnamefont{Fuchs}}, \bibnamefont{and}
  \bibinfo{author}{\bibfnamefont{M.~E.} \bibnamefont{Cates}},
  \bibinfo{journal}{Physical Review E} \textbf{\bibinfo{volume}{67}},
  \bibinfo{pages}{031406} (\bibinfo{year}{2003}).

\bibitem[{\citenamefont{Chen et~al.}(2005)\citenamefont{Chen, Kobelev, and
  Schweizer}}]{Chen2005PRE}
\bibinfo{author}{\bibfnamefont{Y.~L.} \bibnamefont{Chen}},
  \bibinfo{author}{\bibfnamefont{V.}~\bibnamefont{Kobelev}}, \bibnamefont{and}
  \bibinfo{author}{\bibfnamefont{K.~S.} \bibnamefont{Schweizer}},
  \bibinfo{journal}{Physical Review E} \textbf{\bibinfo{volume}{71}},
  \bibinfo{pages}{041405} (\bibinfo{year}{2005}).

\bibitem[{\citenamefont{Puertas et~al.}(2005)\citenamefont{Puertas, Fuchs, and
  Cates}}]{Puertas2005JPCB}
\bibinfo{author}{\bibfnamefont{A.~M.} \bibnamefont{Puertas}},
  \bibinfo{author}{\bibfnamefont{M.}~\bibnamefont{Fuchs}}, \bibnamefont{and}
  \bibinfo{author}{\bibfnamefont{M.~E.} \bibnamefont{Cates}},
  \bibinfo{journal}{Journal of Physical Chemistry B}
  \textbf{\bibinfo{volume}{109}}, \bibinfo{pages}{6666} (\bibinfo{year}{2005}).

\bibitem[{\citenamefont{Groenewold and Kegel}(2004)}]{Groenewold2004JPCM}
\bibinfo{author}{\bibfnamefont{J.}~\bibnamefont{Groenewold}} \bibnamefont{and}
  \bibinfo{author}{\bibfnamefont{W.~K.} \bibnamefont{Kegel}},
  \bibinfo{journal}{Journal of Physics-Condensed Matter}
  \textbf{\bibinfo{volume}{16}}, \bibinfo{pages}{S4877} (\bibinfo{year}{2004}).

\bibitem[{\citenamefont{Dinsmore et~al.}(2001)\citenamefont{Dinsmore, Weeks,
  Prasad, Levitt, and Weitz}}]{Dinsmore2001AO}
\bibinfo{author}{\bibfnamefont{A.~D.} \bibnamefont{Dinsmore}},
  \bibinfo{author}{\bibfnamefont{E.~R.} \bibnamefont{Weeks}},
  \bibinfo{author}{\bibfnamefont{V.}~\bibnamefont{Prasad}},
  \bibinfo{author}{\bibfnamefont{A.~C.} \bibnamefont{Levitt}},
  \bibnamefont{and} \bibinfo{author}{\bibfnamefont{D.~A.} \bibnamefont{Weitz}},
  \bibinfo{journal}{Applied Optics} \textbf{\bibinfo{volume}{40}},
  \bibinfo{pages}{4152} (\bibinfo{year}{2001}).

\bibitem[{\citenamefont{Solomon and Varadan}(2001)}]{Solomon2001PRE}
\bibinfo{author}{\bibfnamefont{M.~J.} \bibnamefont{Solomon}} \bibnamefont{and}
  \bibinfo{author}{\bibfnamefont{P.}~\bibnamefont{Varadan}},
  \bibinfo{journal}{Physical Review E} \textbf{\bibinfo{volume}{63}},
  \bibinfo{pages}{051402} (\bibinfo{year}{2001}).

\bibitem[{\citenamefont{Campbell and Bartlett}(2002)}]{Campbell2002JCIS}
\bibinfo{author}{\bibfnamefont{A.~I.} \bibnamefont{Campbell}} \bibnamefont{and}
  \bibinfo{author}{\bibfnamefont{P.}~\bibnamefont{Bartlett}},
  \bibinfo{journal}{Journal of Colloid and Interface Science}
  \textbf{\bibinfo{volume}{256}}, \bibinfo{pages}{325} (\bibinfo{year}{2002}).

\bibitem[{\citenamefont{Mohraz and Solomon}(2005)}]{Mohraz2005JOR}
\bibinfo{author}{\bibfnamefont{A.}~\bibnamefont{Mohraz}} \bibnamefont{and}
  \bibinfo{author}{\bibfnamefont{M.~J.} \bibnamefont{Solomon}},
  \bibinfo{journal}{Journal of Rheology} \textbf{\bibinfo{volume}{49}},
  \bibinfo{pages}{657} (\bibinfo{year}{2005}).

\bibitem[{\citenamefont{Asakura and Oosawa}(1954)}]{Asakura1954JCP}
\bibinfo{author}{\bibfnamefont{S.}~\bibnamefont{Asakura}} \bibnamefont{and}
  \bibinfo{author}{\bibfnamefont{F.}~\bibnamefont{Oosawa}},
  \bibinfo{journal}{Journal of Chemical Physics} \textbf{\bibinfo{volume}{22}},
  \bibinfo{pages}{1255} (\bibinfo{year}{1954}).

\bibitem[{\citenamefont{Ilett et~al.}(1995)\citenamefont{Ilett, Orrock, Poon,
  and Pusey}}]{Illett1995PRE}
\bibinfo{author}{\bibfnamefont{S.~M.} \bibnamefont{Ilett}},
  \bibinfo{author}{\bibfnamefont{A.}~\bibnamefont{Orrock}},
  \bibinfo{author}{\bibfnamefont{W.~C.~K.} \bibnamefont{Poon}},
  \bibnamefont{and} \bibinfo{author}{\bibfnamefont{P.~N.} \bibnamefont{Pusey}},
  \bibinfo{journal}{Physical Review E} \textbf{\bibinfo{volume}{51}},
  \bibinfo{pages}{1344} (\bibinfo{year}{1995}).

\bibitem[{\citenamefont{Shah et~al.}(2003{\natexlab{b}})\citenamefont{Shah,
  Chen, Schweizer, and Zukoski}}]{Shah2003JCP}
\bibinfo{author}{\bibfnamefont{S.~A.} \bibnamefont{Shah}},
  \bibinfo{author}{\bibfnamefont{Y.~L.} \bibnamefont{Chen}},
  \bibinfo{author}{\bibfnamefont{K.~S.} \bibnamefont{Schweizer}},
  \bibnamefont{and} \bibinfo{author}{\bibfnamefont{C.~F.}
  \bibnamefont{Zukoski}}, \bibinfo{journal}{Journal of Chemical Physics}
  \textbf{\bibinfo{volume}{119}}, \bibinfo{pages}{8747}
  (\bibinfo{year}{2003}{\natexlab{b}}).

\bibitem[{\citenamefont{Solomon and Solomon}(2006)}]{Solomon2006JCP}
\bibinfo{author}{\bibfnamefont{T.}~\bibnamefont{Solomon}} \bibnamefont{and}
  \bibinfo{author}{\bibfnamefont{M.~J.} \bibnamefont{Solomon}},
  \bibinfo{journal}{Journal of Chemical Physics}
  \textbf{\bibinfo{volume}{124}}, \bibinfo{pages}{134905}
  (\bibinfo{year}{2006}).

\bibitem[{\citenamefont{Antl et~al.}(1986)\citenamefont{Antl, Goodwin, Hill,
  Ottewill, Owens, Papworth, and Waters}}]{Antl1986CS}
\bibinfo{author}{\bibfnamefont{L.}~\bibnamefont{Antl}},
  \bibinfo{author}{\bibfnamefont{J.~W.} \bibnamefont{Goodwin}},
  \bibinfo{author}{\bibfnamefont{R.~D.} \bibnamefont{Hill}},
  \bibinfo{author}{\bibfnamefont{R.~H.} \bibnamefont{Ottewill}},
  \bibinfo{author}{\bibfnamefont{S.~M.} \bibnamefont{Owens}},
  \bibinfo{author}{\bibfnamefont{S.}~\bibnamefont{Papworth}}, \bibnamefont{and}
  \bibinfo{author}{\bibfnamefont{J.~A.} \bibnamefont{Waters}},
  \bibinfo{journal}{Colloids and Surfaces} \textbf{\bibinfo{volume}{17}},
  \bibinfo{pages}{67} (\bibinfo{year}{1986}).

\bibitem[{\citenamefont{Royall et~al.}(2003)\citenamefont{Royall, Leunissen,
  and van Blaaderen}}]{Royall2003JPCM}
\bibinfo{author}{\bibfnamefont{C.~P.} \bibnamefont{Royall}},
  \bibinfo{author}{\bibfnamefont{M.~E.} \bibnamefont{Leunissen}},
  \bibnamefont{and} \bibinfo{author}{\bibfnamefont{A.}~\bibnamefont{van
  Blaaderen}}, \bibinfo{journal}{Journal of Physics-Condensed Matter}
  \textbf{\bibinfo{volume}{15}}, \bibinfo{pages}{S3581} (\bibinfo{year}{2003}).

\bibitem[{\citenamefont{Auer et~al.}(2003)\citenamefont{Auer, Poon, and
  Frenkel}}]{Auer2003PRE}
\bibinfo{author}{\bibfnamefont{S.}~\bibnamefont{Auer}},
  \bibinfo{author}{\bibfnamefont{W.~C.~K.} \bibnamefont{Poon}},
  \bibnamefont{and} \bibinfo{author}{\bibfnamefont{D.}~\bibnamefont{Frenkel}},
  \bibinfo{journal}{Physical Review E} \textbf{\bibinfo{volume}{67}},
  \bibinfo{pages}{020401 (R)} (\bibinfo{year}{2003}).

\bibitem[{\citenamefont{Yethiraj and van Blaaderen}(2003)}]{Yethiraj2003N}
\bibinfo{author}{\bibfnamefont{A.}~\bibnamefont{Yethiraj}} \bibnamefont{and}
  \bibinfo{author}{\bibfnamefont{A.}~\bibnamefont{van Blaaderen}},
  \bibinfo{journal}{Nature} \textbf{\bibinfo{volume}{421}},
  \bibinfo{pages}{513} (\bibinfo{year}{2003}).

\bibitem[{\citenamefont{Leunissen et~al.}(2005)\citenamefont{Leunissen,
  Christova, Hynninen, Royall, Campbell, Imhof, Dijkstra, van Roij, and van
  Blaaderen}}]{Leunissen2005N}
\bibinfo{author}{\bibfnamefont{M.~E.} \bibnamefont{Leunissen}},
  \bibinfo{author}{\bibfnamefont{C.~G.} \bibnamefont{Christova}},
  \bibinfo{author}{\bibfnamefont{A.~P.} \bibnamefont{Hynninen}},
  \bibinfo{author}{\bibfnamefont{C.~P.} \bibnamefont{Royall}},
  \bibinfo{author}{\bibfnamefont{A.~I.} \bibnamefont{Campbell}},
  \bibinfo{author}{\bibfnamefont{A.}~\bibnamefont{Imhof}},
  \bibinfo{author}{\bibfnamefont{M.}~\bibnamefont{Dijkstra}},
  \bibinfo{author}{\bibfnamefont{R.}~\bibnamefont{van Roij}}, \bibnamefont{and}
  \bibinfo{author}{\bibfnamefont{A.}~\bibnamefont{van Blaaderen}},
  \bibinfo{journal}{Nature} \textbf{\bibinfo{volume}{437}},
  \bibinfo{pages}{235} (\bibinfo{year}{2005}).

\bibitem[{\citenamefont{Royall et~al.}(2005)\citenamefont{Royall, Aarts, and
  Tanaka}}]{Royall2005JPCM}
\bibinfo{author}{\bibfnamefont{C.~P.} \bibnamefont{Royall}},
  \bibinfo{author}{\bibfnamefont{D.~G.~A.~L.} \bibnamefont{Aarts}},
  \bibnamefont{and} \bibinfo{author}{\bibfnamefont{H.}~\bibnamefont{Tanaka}},
  \bibinfo{journal}{Journal of Physics-Condensed Matter}
  \textbf{\bibinfo{volume}{17}}, \bibinfo{pages}{S3401} (\bibinfo{year}{2005}).

\bibitem[{\citenamefont{Brandrup et~al.}(1999)\citenamefont{Brandrup, Immergut,
  and Grulke}}]{PolymerHandbook}
\bibinfo{author}{\bibfnamefont{J.}~\bibnamefont{Brandrup}},
  \bibinfo{author}{\bibfnamefont{E.~H.} \bibnamefont{Immergut}},
  \bibnamefont{and} \bibinfo{author}{\bibfnamefont{E.~A.}
  \bibnamefont{Grulke}}, \emph{\bibinfo{title}{Polymer Handbook}}
  (\bibinfo{publisher}{Wiley}, \bibinfo{address}{New York},
  \bibinfo{year}{1999}), \bibinfo{edition}{4th} ed.

\bibitem[{\citenamefont{Yamakawa}(1971)}]{YamakawaPolySolns}
\bibinfo{author}{\bibfnamefont{H.}~\bibnamefont{Yamakawa}},
  \emph{\bibinfo{title}{Modern Theory of Polymer Solutions}}, Harper's
  Chemistry Series (\bibinfo{publisher}{Harper and Row},
  \bibinfo{address}{Evanston, IL}, \bibinfo{year}{1971}).

\bibitem[{\citenamefont{Larson}(1998)}]{LarsonComplexFluids}
\bibinfo{author}{\bibfnamefont{R.~G.} \bibnamefont{Larson}},
  \emph{\bibinfo{title}{The Structure and Rheology of Complex Fluids}}
  (\bibinfo{publisher}{Oxford University Press}, \bibinfo{address}{USA},
  \bibinfo{year}{1998}).

\bibitem[{\citenamefont{Chatterjee and Schweizer}(1998)}]{Chatterjee1998JCP}
\bibinfo{author}{\bibfnamefont{A.~P.} \bibnamefont{Chatterjee}}
  \bibnamefont{and} \bibinfo{author}{\bibfnamefont{K.~S.}
  \bibnamefont{Schweizer}}, \bibinfo{journal}{Journal of Chemical Physics}
  \textbf{\bibinfo{volume}{109}}, \bibinfo{pages}{10464}
  (\bibinfo{year}{1998}).

\bibitem[{\citenamefont{Crocker and Grier}(1996)}]{Crocker1996JCIS}
\bibinfo{author}{\bibfnamefont{J.~C.} \bibnamefont{Crocker}} \bibnamefont{and}
  \bibinfo{author}{\bibfnamefont{D.~G.} \bibnamefont{Grier}},
  \bibinfo{journal}{Journal Colloid Interface Science}
  \textbf{\bibinfo{volume}{179}}, \bibinfo{pages}{298} (\bibinfo{year}{1996}).

\bibitem[{\citenamefont{Savin and Doyle}(2005)}]{Savin2005BJ}
\bibinfo{author}{\bibfnamefont{T.}~\bibnamefont{Savin}} \bibnamefont{and}
  \bibinfo{author}{\bibfnamefont{P.~S.} \bibnamefont{Doyle}},
  \bibinfo{journal}{Biophysical Journal} \textbf{\bibinfo{volume}{88}},
  \bibinfo{pages}{623} (\bibinfo{year}{2005}).

\bibitem[{\citenamefont{Hansen and McDonald}(1990)}]{HansenSimpleLiquids}
\bibinfo{author}{\bibfnamefont{J.~P.} \bibnamefont{Hansen}} \bibnamefont{and}
  \bibinfo{author}{\bibfnamefont{I.~R.} \bibnamefont{McDonald}},
  \emph{\bibinfo{title}{Theory of Simple Liquids}}
  (\bibinfo{publisher}{Academic Press}, \bibinfo{address}{London; New York},
  \bibinfo{year}{1990}), \bibinfo{edition}{2nd} ed.

\bibitem[{\citenamefont{Kob et~al.}(1997)\citenamefont{Kob, Donati, Plimpton,
  Poole, and Glotzer}}]{Kob1997PRL}
\bibinfo{author}{\bibfnamefont{W.}~\bibnamefont{Kob}},
  \bibinfo{author}{\bibfnamefont{C.}~\bibnamefont{Donati}},
  \bibinfo{author}{\bibfnamefont{S.~J.} \bibnamefont{Plimpton}},
  \bibinfo{author}{\bibfnamefont{P.~H.} \bibnamefont{Poole}}, \bibnamefont{and}
  \bibinfo{author}{\bibfnamefont{S.~C.} \bibnamefont{Glotzer}},
  \bibinfo{journal}{Physical Review Letters} \textbf{\bibinfo{volume}{79}},
  \bibinfo{pages}{2827} (\bibinfo{year}{1997}).

\bibitem[{\citenamefont{Russel et~al.}(1991)\citenamefont{Russel, Saville, and
  Schowalter}}]{RusselColloidalDispersions}
\bibinfo{author}{\bibfnamefont{W.~B.} \bibnamefont{Russel}},
  \bibinfo{author}{\bibfnamefont{D.~A.} \bibnamefont{Saville}},
  \bibnamefont{and} \bibinfo{author}{\bibfnamefont{W.~R.}
  \bibnamefont{Schowalter}}, \emph{\bibinfo{title}{Colloidal Dispersions}}
  (\bibinfo{publisher}{Cambridge University Press}, \bibinfo{address}{New
  York}, \bibinfo{year}{1991}).

\bibitem[{\citenamefont{Lionberger and Russel}(1997)}]{Lionberger1997JOR}
\bibinfo{author}{\bibfnamefont{R.~A.} \bibnamefont{Lionberger}}
  \bibnamefont{and} \bibinfo{author}{\bibfnamefont{W.~B.}
  \bibnamefont{Russel}}, \bibinfo{journal}{Journal of Rheology}
  \textbf{\bibinfo{volume}{41}}, \bibinfo{pages}{399} (\bibinfo{year}{1997}).

\bibitem[{\citenamefont{Kroy et~al.}(2004)\citenamefont{Kroy, Cates, and
  Poon}}]{Kroy2004PRL}
\bibinfo{author}{\bibfnamefont{K.}~\bibnamefont{Kroy}},
  \bibinfo{author}{\bibfnamefont{M.~E.} \bibnamefont{Cates}}, \bibnamefont{and}
  \bibinfo{author}{\bibfnamefont{W.~C.~K.} \bibnamefont{Poon}},
  \bibinfo{journal}{Phys. Rev. Lett.} \textbf{\bibinfo{volume}{92}},
  \bibinfo{pages}{148302 1} (\bibinfo{year}{2004}).

\bibitem[{\citenamefont{Ramakrishnan et~al.}(2004)\citenamefont{Ramakrishnan,
  Chen, Schweizer, and Zukoski}}]{Ramakrishnan2004PRE}
\bibinfo{author}{\bibfnamefont{S.}~\bibnamefont{Ramakrishnan}},
  \bibinfo{author}{\bibfnamefont{Y.~L.} \bibnamefont{Chen}},
  \bibinfo{author}{\bibfnamefont{K.~S.} \bibnamefont{Schweizer}},
  \bibnamefont{and} \bibinfo{author}{\bibfnamefont{C.~F.}
  \bibnamefont{Zukoski}}, \bibinfo{journal}{Physical Review E}
  \textbf{\bibinfo{volume}{70}}, \bibinfo{pages}{040401 (R)}
  (\bibinfo{year}{2004}).

\bibitem[{\citenamefont{Bergenholtz et~al.}(2003)\citenamefont{Bergenholtz,
  Poon, and Fuchs}}]{Bergenholtz2003Lang}
\bibinfo{author}{\bibfnamefont{J.}~\bibnamefont{Bergenholtz}},
  \bibinfo{author}{\bibfnamefont{W.~C.~K.} \bibnamefont{Poon}},
  \bibnamefont{and} \bibinfo{author}{\bibfnamefont{M.}~\bibnamefont{Fuchs}},
  \bibinfo{journal}{Langmuir} \textbf{\bibinfo{volume}{19}},
  \bibinfo{pages}{4493} (\bibinfo{year}{2003}).

\bibitem[{\citenamefont{Kumar and Douglas}(2001)}]{Kumar2001PRL}
\bibinfo{author}{\bibfnamefont{S.~K.} \bibnamefont{Kumar}} \bibnamefont{and}
  \bibinfo{author}{\bibfnamefont{J.~F.} \bibnamefont{Douglas}},
  \bibinfo{journal}{Physical Review Letters} \textbf{\bibinfo{volume}{87}},
  \bibinfo{pages}{188301} (\bibinfo{year}{2001}).

\bibitem[{\citenamefont{Zaccarelli et~al.}(2005)\citenamefont{Zaccarelli,
  Buldyrev, La~Nave, Moreno, Saika-Voivod, Sciortino, and
  Tartaglia}}]{Zaccarelli2005PRL}
\bibinfo{author}{\bibfnamefont{E.}~\bibnamefont{Zaccarelli}},
  \bibinfo{author}{\bibfnamefont{S.~V.} \bibnamefont{Buldyrev}},
  \bibinfo{author}{\bibfnamefont{E.}~\bibnamefont{La~Nave}},
  \bibinfo{author}{\bibfnamefont{A.~J.} \bibnamefont{Moreno}},
  \bibinfo{author}{\bibfnamefont{I.}~\bibnamefont{Saika-Voivod}},
  \bibinfo{author}{\bibfnamefont{F.}~\bibnamefont{Sciortino}},
  \bibnamefont{and}
  \bibinfo{author}{\bibfnamefont{P.}~\bibnamefont{Tartaglia}},
  \bibinfo{journal}{Physical Review Letters} \textbf{\bibinfo{volume}{94}},
  \bibinfo{pages}{218301} (\bibinfo{year}{2005}).

\end{thebibliography}
\end{document}